\documentclass[10pt]{article}
\usepackage{charter} 
\usepackage{fullpage}
\usepackage{cite}
\usepackage{amsmath,amssymb,amsfonts}
\usepackage{graphicx}
\usepackage{textcomp}
\usepackage{booktabs}

\usepackage{multirow}
\usepackage[flushleft]{threeparttable}
\usepackage{subcaption}
\usepackage{bm}
\usepackage{algorithm}
\usepackage{algpseudocode}
\usepackage{hyperref}
\usepackage[font=small,skip=2pt]{caption}

\usepackage{enumitem}

\newcommand{\markup}[1]{{{#1}}}

\def\argmin{\mathop{\mathrm{arg\,min}}} 

\def\lim{\mathop{\mathrm{lim}}} 

\newcommand{\norm}[1]{\left\lVert#1\right\rVert}

\def\ebm{{\bm{e}}}

\def\xbm{{\bm{x}}}
\def\gbm{{\bm{g}}}
\def\ybm{{\bm{y}}}

\def\rbm{{\bm{r}}}

\def\vbm{{\bm{v}}}

\def\varphibm{{\bm{\varphi}}}

\def\thetabm{{\bm{\theta}}}
\def\phibm{{\bm{\phi}}}

\def\Hbm{{\bm{H}}}
\def\Dbm{{\bm{D}}}
\def\Fbm{{\bm{F}}}
\def\Sbm{{\bm{S}}}
\def\Pbm{{\bm{P}}}

\def\Ibm{{\bm{I}}}

\def\xbmhat{{\widehat{\bm{x}}}}
\def\vbmhat{{\widehat{\bm{v}}}}

\def\Tsf{{\mathrm{T}}}

\def\C{\mathbb{C}}
\def\R{\mathbb{R}}

\def\Dcal{{\mathcal{D}}}
\def\Rcal{{\mathcal{R}}}

\def\phibm{{\bm{\phi}}}
\def\phibmhat{{\bm{\hat{\phi}}}}

\def\mbm{{\bm{m}}}

\begin{document}

\title{\markup{Deformation-Compensated Learning for Image Reconstruction without Ground Truth}}

\author{
Weijie Gan\thanks{Weijie Gan and Yu Sun are with the Department of Computer Science \& Engineering, Washington University in St. Louis, St. Louis, MO 63130 USA (e-mail: weijie.gan@wustl.edu; sun.yu@wustl.edu).} , 
Yu Sun\footnotemark[1] , 
Cihat Eldeniz\thanks{Cihat Eldeniz is with the Mallinckrodt Institute of Radiology, Washington University in St. Louis, St. Louis, MO 63130 USA (e-mail: cihat.eldeniz@ wustl.edu).} ,
Jiaming Liu\thanks{Jiaming Liu is with the Department of Electrical \& System Engineering, Washington University in St. Louis, St. Louis, MO 63130 USA (e-mail: jiaming.liu@wustl.edu).} , 
Hongyu An\thanks{Hongyu An is with the Mallinckrodt Institute of Radiology, Department of Neurology, Department of Biomedical Engineering, Saint Louis, MO 63130 USA, and also with the Division of Biology and Biomedical Sciences, Washington University in St. Louis, St. Louis, MO 63130 USA (e-mail: hongyuan@wustl.edu).} ,
Ulugbek S. Kamilov\thanks{Ulugbek S. Kamilov is with the Department of Computer Science \& Engineering and Electrical \& Systems Engineering, Washington University in St. Louis, St. Louis, MO 63130 USA (e-mail: kamilov@ieee.org).}
\vspace{-2\baselineskip}
}

\date{}
\maketitle

\begin{abstract}
Deep neural networks for medical image reconstruction are traditionally trained using high-quality ground-truth images as training targets. Recent work on \emph{Noise2Noise (N2N)} has shown the potential of using multiple noisy measurements of the same object as an alternative to having a \markup{ground-truth}. \markup{However, existing N2N-based methods are not suitable for learning from the measurements of an object undergoing nonrigid deformation. This paper addresses this issue by proposing the \emph{deformation-compensated learning (DeCoLearn)} method for training deep reconstruction networks by compensating for object deformations. A key component of DeCoLearn is a deep registration module, which is jointly trained with the deep reconstruction network without any ground-truth supervision.} We validate \markup{DeCoLearn} on both simulated and experimentally collected \emph{magnetic resonance imaging (MRI)} data and show that it significantly improves imaging quality.
\end{abstract}


\begin{figure}
    \centering
    \includegraphics[width=.415\textwidth]{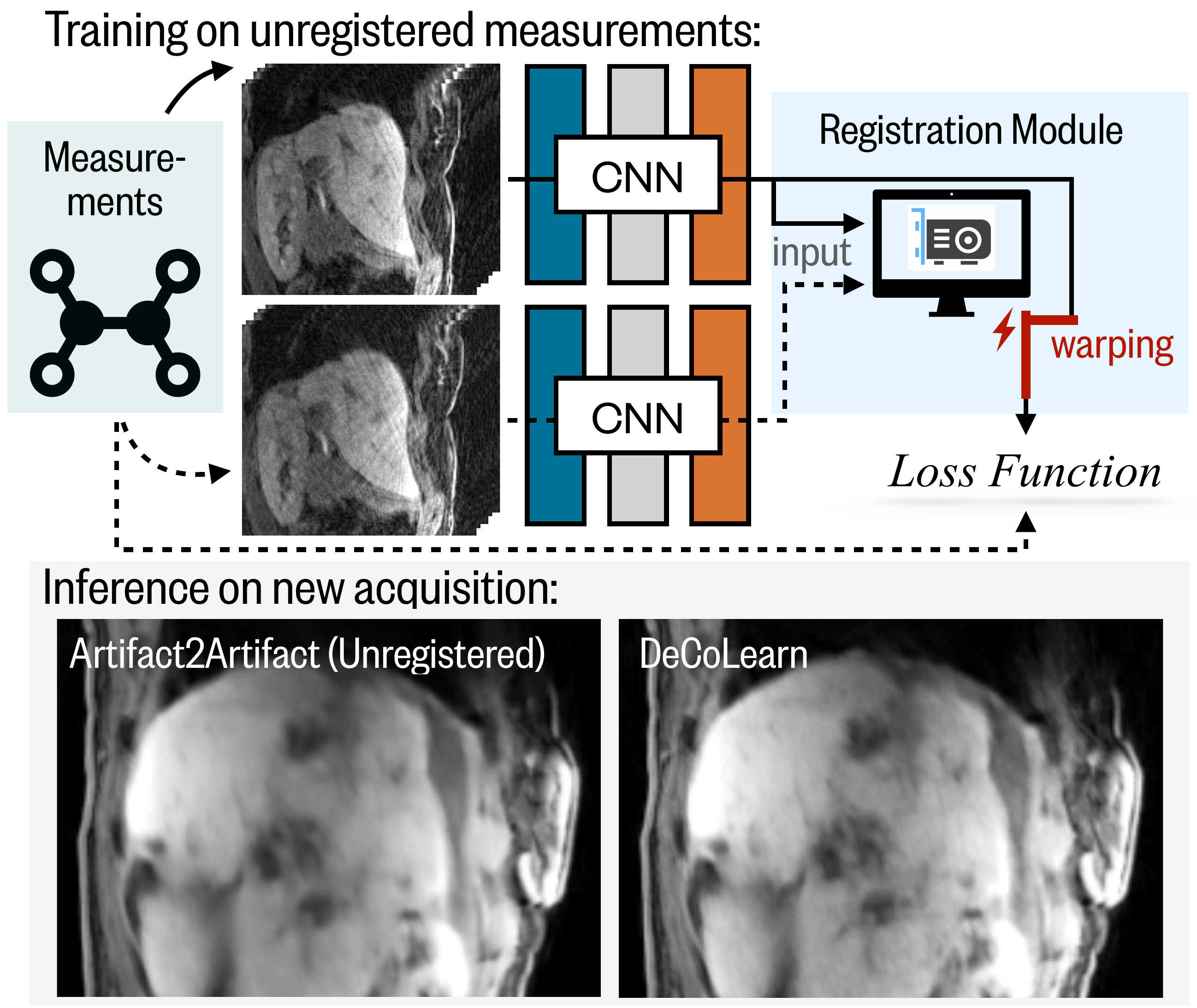}
    \caption{ 
    The conceptual illustration of \markup{DeCoLearn} for CS-MRI~\cite{lustigSparse2007}. 
    \markup{DeCoLearn} trains a \emph{convolutional neural network (CNN)} on unregistered measurements using a registration module that corrects for object \markup{deformation}.
    \markup{This example highlights the improvement of \markup{DeCoLearn} over an identical deep reconstruction network trained on the same measurements but without deformation compensation.}
    } 
    \label{fig:intro}
\end{figure}

\section{Introduction}
\label{sec:introduction}
{T}{he} recovery of a high-quality image from a set of noisy measurements is fundamental in medical imaging. 
For instance, it is essential in \emph{compressed sensing magnetic resonance imaging (CS-MRI)}~\cite{lustigSparse2007}, which aims at obtaining diagnostic-quality images from severely undersampled k-space measurements. 
The recovery is traditionally formulated as an \emph{inverse problem} that leverages a forward model characterizing the physics of data acquisition and a regularizer imposing prior knowledge on the solution.
Many regularizers have been proposed to date, including those based on transform-domain sparsity, low-rank penalty, and dictionary learning~\cite{danielyanBM3D2011, eladImage2006, huFast2011, rudinNonlinear1992}.

\emph{Deep learning (DL)} has recently gained popularity in medical image reconstruction~\cite{knollDeeplearning2020, lucasUsing2018, mccannConvolutional2017, ongieDeep2020, wangDeep2020}.
A \markup{widely-used} DL strategy is based on training a \emph{convolutional neural network (CNN)} to map a low-quality image to its desired high-quality counterpart. However, this simple supervised DL approach is impractical in applications where it is difficult to collect a sufficient number of high-quality training images. This limitation has motivated the research on ``ground-truth-free'' DL schemes that rely exclusively on the information available in the corrupted data itself~\cite{akccakaya2021unsupervised, lehtinenNoise2Noise2018, liuRARE2020, krullNoise2voidlearning2019, yamanSelfsupervised2020}. In this study, we focus on the line of work based on \emph{{Noise2Noise (N2N)}~\cite{lehtinenNoise2Noise2018}}, which has shown that one can train a CNN without \markup{ground-truth} by using only pairs of noisy observations of the same object. Recent extensions to N2N have investigated the potential of this strategy in a variety of imaging scenarios~\cite{laineHighquality2019, ganPhase2Phaseinpress, torop2020deep, ehret2019model,yu2020joint, jiang2020weakly, buchholzCryoCARE2019, hendriksenNoise2Inverse2020, xu2021deformed2self}.

Despite recent progress, current N2N-based methods inherently assume that the object is stationary across all the measurements. \markup{This assumption limits their ability to exploit measurements of an object undergoing nonrigid deformation}. To overcome this limitation, we propose \markup{a new \emph{deformation-compensated learning (DeCoLearn})} method that uses multiple measurements of a \markup{deformation}-affected object by integrating a deep registration~\cite{fuDeep2020} module into the deep architecture for an end-to-end training. \markup{DeCoLearn} enables training without any ground-truth supervision by adopting recent ideas from self-supervised deep registration~\cite{devosDeep2019,lei4dct2019,yooSsEMnet2017,balakrishnanVoxelMorph2019}. \markup{The key contributions of this work are as follows:}
\begin{itemize}
    \item \markup{DeCoLearn extends N2N and its more recent variant \emph{Artifact2Artifact (A2A)}~\cite{liuRARE2020} to enable learning directly in the measurement domain (e.g., k-space for MRI) from undersampled and noisy measurements without any fully sampled \markup{ground-truth}. It is trained by transforming the reconstructed images back to the measurement domain and minimizing the difference between the predicted measurements and the measured raw data.}
    
    \item \markup{DeCoLearn can use information from multiple measurements of an object undergoing nonrigid deformation, which enables it to leverage information that is not suitable for direct N2N/A2A training.} This capability is achieved by integrating a deep registration module into the final architecture (see Fig.~\ref{fig:method}), which is trained end-to-end on unregistered, noisy, and subsampled measurements. Note that the registration module is only necessary during training, since image reconstruction can be performed by using only the reconstruction module.

    \item \markup{We extensively validate \markup{DeCoLearn} on both simulated and experimentally collected MRI data.} Our simulation results show that \markup{DeCoLearn} quantitatively \markup{outperforms several baseline methods} and matches the performance of \markup{oracle method that has the knowledge of the true object motion}. Our results on experimentally collected data show that \markup{DeCoLearn} leads to significant quality improvements by using additional measurements \markup{not suitable for traditional N2N-based learning}.

\end{itemize}

This paper extends the preliminary work~\cite{ganDeep2020} presented at the 2021 IEEE International Symposium on Biomedical Imaging by including additional technical details\markup{, comparison against several state-of-the-art methods,} and validation on experimentally collected MRI data.

\section{Background}
\label{sec-material}

\subsection{Imaging Inverse Problems}
We consider the problem of recovering an unknown image $\xbm \in \C^n$ from its noisy measurements $\ybm\in\C^m$ specified by the linear system
\begin{equation}
    \label{equ:imaging}
    \ybm = \Hbm\xbm + \ebm\ ,
\end{equation}
where $\ebm\in\C^m$ is noise and $\Hbm\in\C^{m\times n}$ is the measurement operator that characterizes the response of the imaging system.
For instance, $\Hbm$ in parallel CS-MRI with a dynamic object can be represented as
\begin{equation}
    \label{equ:mcnufft}
    \Hbm_i^{(t)} = \Pbm^{(t)}\Fbm\Sbm_i\ ,
\end{equation}
where $\Fbm$ denotes the Fourier transform operator, $\Pbm^{(t)}$ refers to a k-space sampling operator at time $t$, and $\Sbm_i$ is the matrix of the pixel-wise sensitivity map of the $i$th coil. 
\markup{We assume that $\Sbm_i$ is fixed over time.}
\markup{When $m < n$, the problem is an ill-posed inverse problem, which can be conventionally formulated as regularized optimization}
\begin{equation}
    \label{equ:optimization}
    \argmin_{\xbm\in\C^n}\ \markup{\Dcal}(\xbm) + \markup{\Rcal}(\xbm)\ ,
\end{equation}
where $\markup{\Dcal}$ is the data-fidelity term that quantifies consistency with the observed data $\ybm$ and $\markup{\Rcal}$ is a regularizer that encodes prior knowledge on $\xbm$. For example, two \markup{widely-used} functions in imaging are the least-squares and total variation (TV)
\begin{equation}
    \label{equ:optimization_tv}
    \markup{\Dcal}(\xbm) = \frac{1}{2}\norm{\Hbm\xbm-\ybm}^2_2\quad \mathrm{and}\quad \markup{\Rcal}(\xbm)=\tau\norm{\Dbm\xbm}_1\ ,
\end{equation}
where $\tau > 0$ controls the regularization strength and $\Dbm$ is the discrete gradient operator~\cite{rudinNonlinear1992}.


In the past few years, DL has gained popularity for solving imaging inverse problems due to its excellent performance (see reviews in~\cite{knollDeeplearning2020, lucasUsing2018, mccannConvolutional2017, ongieDeep2020, wangDeep2020}). One widely-used DL approach is based on training a CNN $\mathrm{h}_\thetabm(\cdot)$, with parameters $\thetabm\in\R^p$,  to compute a regularized inverse of $\Hbm$ by mapping corrupted images to their clean target versions. The training can be formulated as an optimization problem
\begin{equation}
    \argmin_\thetabm \sum_i \mathcal{L}(\mathrm{h}_\thetabm(\Hbm^\dagger_i\ybm_i), \xbm_i)\ ,
\end{equation}
where $\Hbm^\dagger$ is a \markup{pseudoinverse} of $\Hbm$, $\mathcal{L}$ is a loss function, and $i$ indexes the samples in the training set. Popular choices for $\mathcal{L}$ include the $\ell_1$ and $\ell_2$ norms.
For example, prior work on DL for CS-MRI has trained the CNN by mapping the zero-filled images to their corresponding fully-sampled ground-truth images~\cite{aggarwalMoDL2019, schlemperDeep2018, yangDeep2016}. While traditional DL relies on generic CNN architectures (such as UNet~\cite{ronnebergerUnet2015}), recent work has also explored the integration of DL and model-based optimization. For example, \emph{plug-and-play priors  (PnP)}~\cite{venkatakrishnan2013plug} and \emph{regularization by denoisers (RED)}~\cite{romano2017little} refer to a related family of algorithms that use pre-trained deep denoisers as imaging priors~\cite{sreehari2016plug, zhang2017learning, sun2019regularized, zhang2019deep}. The recent publication~\cite{ahmad2020plug} has reviewed PnP/RED in the context of image reconstruction for MRI. \emph{Deep unrolling} is another widely-used strategy inspired by LISTA~\cite{gregor2010learning}, where the iterations of a regularized optimization are interpreted as layers of a CNN and trained in an end-to-end fashion~\cite{yangDeep2016, zhang2018ista, aggarwalMoDL2019, chen2015learning, schlemperDeep2018, gregor2010learning, liu2021sgd}.

Our work contributes to this broad area by providing a new DL method that does not require clean ground-truth images as training targets. While this work focuses on traditional model-free DL architectures, our method is fully compatible with the latest model-based architectures.

\subsection{\markup{Deep Image Reconstruction without Ground Truth}}
\label{sec-background-gtfree} 

There is a growing interest in DL image reconstruction to reduce the dependence on high-quality ground-truth training targets. 
One widely-adopted framework is \emph{N2N}~\cite{lehtinenNoise2Noise2018}, where the CNN $\mathrm{h}_\thetabm$ is trained on a group of noisy images $\{\xbmhat_{ij}\}$, with $j$ indexing different realizations of the same underlying image $i$.
There have been multiple extensions of the original method~\cite{laineHighquality2019, ganPhase2Phaseinpress, torop2020deep, ehret2019model, yu2020joint, jiang2020weakly, buchholzCryoCARE2019, hendriksenNoise2Inverse2020, xu2021deformed2self} with applications to numerous medical imaging problems, including motion-resolved MRI~\cite{liuRARE2020, ganPhase2Phaseinpress}, cryo-transmission electron microscopy (cryo-TEM)~\cite{buchholzCryoCARE2019} and optical coherence tomography angiography (OCTA)~\cite{jiang2020weakly}.
\emph{A2A}~\cite{liuRARE2020} is one of the extensions of N2N that showed excellent performance using multiple noisy and artifact-corrupted images $\{\xbmhat_{ij}\}$ obtained directly from sparsely-sampled MR measurements. In A2A, $ij$ denotes the $j^{\text{th}}$ MRI acquisition of the subject $i$ with each acquisition consisting \markup{a different} undersampling pattern and noise realization. 
\markup{The whole dataset $\{\xbmhat_{ij}\}$ is assumed to compliment the information missing in each individual measurement, therefore enabling training of the CNN $\mathrm{h}_\thetabm$ to predict clean images. The underlying assumption of N2N/A2A is that the expected value of the images $\{\xbmhat_{ij}\}_j$ still matches the ground-truth $\xbm_i$~\cite{lehtinenNoise2Noise2018}}. 
The CNN \markup{in A2A} is trained by minimizing a loss function
\begin{equation}
    \label{n2n}
    \argmin_\thetabm \sum_{i,j,j'} \mathcal{L}\big(\mathrm{h}_\thetabm(\xbmhat_{ij}), \ \xbmhat_{ij'}\big)\ .
\end{equation} 
Recent works~\cite{yaman2020self:mrm,yamanSelfsupervised2020} have \markup{shown} the potential of training a model-based deep network without \markup{ground-truth} by dividing a single k-space MRI acquisition into two subsets and using both subsampled sets of measurements as training targets.
\markup{The same training strategy has been extended to the ``zero-shot'' learning and achieved excellent performance when training and testing datasets are highly inconsistent~\cite{yaman2021zeroshot}.}
A similar strategy has also been used for denoising in 3D parallel-beam tomography by splitting a stack of noisy sinograms along the angular axis~\cite{hendriksenNoise2Inverse2020}.
Two recent papers considered the inclusion of image deformation into the training of a deep image denoiser~\cite{yu2020joint, xu2021deformed2self}. In~\cite{yu2020joint}, a pre-trained registration network is used for training a video denoising network. In~\cite{xu2021deformed2self}, a deep network is trained along with a deep deformation network to remove common types of noise in medical images, including additive white Gaussian noise (AWGN), Rician noise, and Poisson noise. The key difference of our work is that it goes beyond denoising by considering general inverse problems and using training labels directly in the k-space for MRI.

\markup{\emph{Noise2Void}~\cite{krullNoise2voidlearning2019} and \emph{Noise2Self}~\cite{batsonNoise2self2019} are a related class of methods that use a single noisy copy of each training image in the dataset~\cite{krullProbabilistic2020, soltanayevTraining2018}}. 
Self2Self~\cite{quanSelf2Self2020} extends this idea to use only a single noisy image as a training sample. These methods have been shown to achieve excellent performance in the context of image denoising. \markup{Since \markup{N2V}-type methods learn only from a single image, they are expected to be suboptimal when dealing with structured artifacts, such as aliasing or streaks. We empirically verify this limitation of N2V in the context of accelerated MRI in Section~\ref{sec:exp}.}

\markup{
Another related line of work is on \emph{deep image prior (DIP)}~\cite{ulyanov2018deep}, where a CNN is used for image reconstruction without any training on external data~\cite{liu2019image, yoo2021time, mataev2019deepred}.
DIP exploits the architecture of the CNN to regularize the reconstruction by mapping random but fixed latent inputs to noisy measurements. A recent method TDDIP~\cite{yoo2021time} extends DIP to dynamic MRI by compensating for the object motion by encoding the motion trajectory into the input latent variable. DIP is fundamentally different from DeCoLearn since it is not an end-to-end DL model and needs to solve a nonconvex optimization problem for each reconstruction task.
} 

Our work contributes to this area by enabling the use of information from the measurements of \markup{an object undergoing nonrigid deformation}. It not only allows our method to use more information for training,  but also addresses the assumptions of stationarity and artifact incoherence in the prior work. \markup{It is worth mentioning that while in this paper we use a traditional CNN as the deep reconstruction network for DeCoLearn, the method itself is fully compatible with any model-based DL architectures~\cite{yamanSelfsupervised2020}.}

\subsection{Deep Image Registration}
\label{sec-background-registration}
Let $\rbm$ and $\mbm$ denote a reference image and its deformed counterpart, respectively. 
Deformable image registration aims to obtain a registration field $\phibmhat^{m \rightarrow r}$ that maps the coordinates of $\mbm$ to those of $\rbm$ by comparing the content of the corresponding images. Deformable image registration has been \markup{widely-used} in many applications, such as motion tracking~\cite{yang2012nonrigid} and image segmentation~\cite{han2008atlas, fu2017automatic}.
The registration field $\phibmhat^{m \rightarrow r}$ is often characterized by a displacement vector field $\vbmhat^{m \rightarrow r}$ that represents coordinate offsets from $\mbm$ to $\rbm$, $\phibmhat^{m \rightarrow r} = \Ibm + \vbmhat^{m \rightarrow r}$, where $\Ibm$ denotes an identity transformation~\cite{bajcsyMultiresolution1989}.

Recently, there has been considerable interest in developing DL methods for deformable image registration~\cite{fuDeep2020},
especially methods that require no knowledge of the ground-truth transformation for training~\cite{balakrishnanVoxelMorph2019, devosDeep2019, lei4dct2019, yooSsEMnet2017}.
The corresponding self-supervised methods train a CNN $\mathrm{g}_\varphibm$, with parameters $\varphibm \in \R^k$, by mapping an input image pair $\{\mbm, \rbm\}$ to a deformation field $\phibmhat^{m \rightarrow r}=\mathrm{g}_\varphibm(\mbm, \rbm)$ that can be used for registration~\cite{fuDeep2020}. 
The CNN is trained on a set of image pairs $\big\{\mbm_i, \rbm_i\big\}$ by minimizing the following loss function
\begin{equation}
    \label{VoxelMorph}
    \argmin_\varphibm \sum_i \mathcal{L}_\mathrm{d}(\mbm_i \circ \phibmhat^{m \rightarrow r}_i, \rbm_i) + \mathcal{L}_\mathrm{r}(\phibmhat^{m \rightarrow r}_i)\ ,
\end{equation}
where $\circ$ is the warping operator that transforms the coordinates of $\mbm_i$ based on the registration field $\phibmhat^{m \rightarrow r}_i$. The term $\mathcal{L}_{\mathrm{d}}$ penalizes the discrepancy between $\mbm_i$ after transformation and its reference $\rbm_i$, while $\mathcal{L}_{\mathrm{r}}$ regularizes the local spatial variations in the estimated registration field.
In order to use the standard gradient methods for minimizing this loss function, the warping operator needs to be differentiable and is often implemented as the \emph{Spatial Transform Network (STN)}~\cite{jaderbergSpatial2015}.

Our work seeks to leverage the recent progress in deep image registration to enable a novel methodology for training deep reconstruction networks on \markup{deformation}-affected datasets.

\begin{figure*}
    \centering
    \includegraphics[width=.935\textwidth]{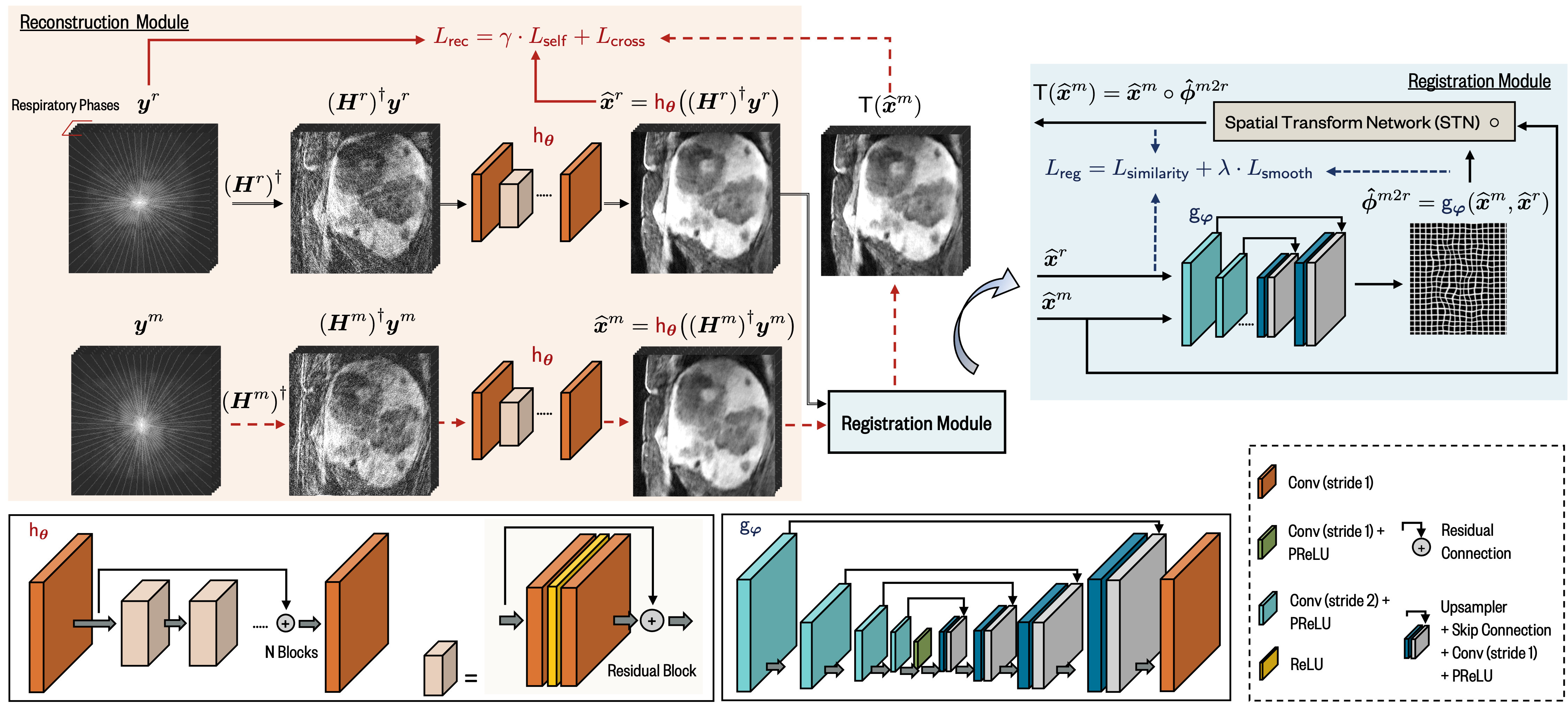}
    \caption{The proposed method jointly trains two CNN modules: $\mathrm{h}_\thetabm$ for image reconstruction and $\mathrm{g}_\varphibm$ for image registration. Inputs are the measurement pairs of the same object but at different motion states. The zero-filled images are passed through $\mathrm{h}_\thetabm$ to remove artifacts due to noise and undersampling. The output images are then used in $\mathrm{g}_\varphibm$ to obtain the motion field characterizing the directional mapping between their coordinates. We implement the warping operator as the \emph{Spatial Transform Network (STN)} to register one of the reconstructed images to the other. We train the whole network end-to-end without any ground-truth images or transformations.}
    \label{fig:method}
\end{figure*}

\subsection{\markup{Motion-Compensated Reconstruction}} 

\markup{\emph{Motion-compensated (MoCo)} reconstruction refers to a class of methods for reconstructing dynamic object from their noisy measurements~\cite{fengXD2016,fengHighly2013, otazoCombination2010,usmanMotion2013,cruzHighly2017,bustin3D2020,blumeJoint2010, odilleJoint2016,coronaMultitasking2019,munozSelfsupervised2022,qiEnd2021}. MoCo methods seek to leverage data redundancy over the motion dimension during reconstruction. For example, traditional model-based MoCo methods include an additional regularizer in the motion dimension~\cite{fengXD2016,fengHighly2013, otazoCombination2010} or enforce spatial smoothness in the images at different motion phases using \emph{motion vector fields (MVFs)}~\cite{usmanMotion2013,cruzHighly2017,bustin3D2020}.
MVFs can be obtained by registering images of the reconstructed object at different motion states or via joint optimization using multi-task optimization~\cite{blumeJoint2010, odilleJoint2016,coronaMultitasking2019}. 
Recent methods have also used DL to estimate MVFs by training a self-supervised network on reconstructed images~\cite{munozSelfsupervised2022} or by jointly updating both MVFs and images in a supervised fashion~\cite{qiEnd2021}.}

\markup{DeCoLearn is a complementary paradigm to the traditional MoCo image reconstruction. The primary focus of DeCoLearn is to enable \emph{learning} given pairs of measurements of objects undergoing deformations. Thus, unlike MoCo methods, DeCoLearn does not specifically target \emph{sequential} data. DeCoLearn can be used both as a traditional (non-MoCo) algorithm on 2D/3D spatial images or extended to explicitly take into account the motion/temporal dimension of the signal.}

\begin{algorithm}[t]
    \caption{\textrm{\markup{DeCoLearn training}} 
    }
    \label{algo} 
    \begin{algorithmic}[1]
    \Require Initial parameters $\thetabm^0 \text{and }\varphibm^0$, number of iterations $K$, and Adam~\cite{kingmaAdam2014} optimizers $\mathrm{Adam}_\mathrm{reg}$ and $\mathrm{Adam}_\mathrm{rec}$.
    \For{number of training iterations $k=1,2,...,K$}
        \State Select a training mini-batch: $\ybm^{r}_i,\ybm^{m}_i,\Hbm^{r}_i,\Hbm^{m}_i$ 

        \State $\thetabm^k \leftarrow \mathrm{Adam}_\mathrm{rec}(\thetabm^{k-1}, \partial L_\mathrm{rec}/\partial\thetabm)$

        \State $\varphibm^k \leftarrow \mathrm{Adam}_\mathrm{reg}(\varphibm^{k-1}, \partial L_\mathrm{reg}/\partial\varphibm)$
    \EndFor

    \State \Return Learned parameters $\thetabm^K$ and $\varphibm^K$.
    \end{algorithmic}
\end{algorithm}

\section{Proposed Method}
\label{sec-method} 
In this section, we introduce the technical details of the proposed method. We start by describing the overall architecture, followed by the details of each module.

\subsection{Overall Model}

Consider a pair of unregistered measurements $(\ybm^r,\ybm^m)$ obtained separately from the same object
\begin{subequations}
\label{equ:pro-fwd}
\begin{align}
    \label{equ:pro-fwd-r}
    \ybm^r &= \Hbm^r \xbm^r + \ebm^r\ \text{and} \\
    \label{equ:pro-fwd-m}
    \ybm^m &= \Hbm^m \xbm^m + \ebm^m\ \text{with}\  \xbm^m = \xbm^r \circ \phibm^{r\rightarrow m} \ ,
\end{align}
\end{subequations}
where $(\Hbm^r, \Hbm^m)$ and $(\ebm^r, \ebm^m)$ denote distinct forward operators and noise vectors, respectively.
Eq.~\eqref{equ:pro-fwd-m} models the object motion as a dense nonrigid transformation-field $\phibm^{r\rightarrow m}$ relative to $\xbm^r$.
For example, $(\ybm^r,\ybm^m)$ can be two motion-affected accelerated MRI measurements of the same patient.
Our method aims to train a deep neural network on a set of such pairs $\{(\ybm^r_i,\ybm^m_i)\}_i^N$, where $N \geq 1$ denotes the total number of training samples, without the need for ground-truth images ($\xbm^r_i$ and $\xbm^m_i$) or transformations ($\phibm^{r\rightarrow m}_i$).

Fig.~\ref{fig:method} summarizes the data processing pipeline of \markup{DeCoLearn}.
It consists of a reconstruction module trained to form images from measurements, and a registration module for registering the reconstructed images onto each other.
The trainable parameters of both modules are denoted as $\thetabm$ and $\varphibm$ in respective order. 
During training, we define two distinct loss functions $\mathcal{L}_\mathrm{rec}$ and $\mathcal{L}_\mathrm{reg}$ as well as two Adam~\cite{kingmaAdam2014} optimizers $\mathrm{Adam}_\mathrm{rec}$ and $\mathrm{Adam}_\mathrm{reg}$ for each module.
Given a mini-batch of training samples, the proposed training procedure alternatively minimizes the loss functions by fixing the trainable parameters of one module while training the other.  Algorithm \ref{algo} summarizes the training strategy.
Note that the registration module of \markup{DeCoLearn} is only employed during training, since reconstruction during testing can be performed directly by using the reconstruction module alone.

\subsection{Reconstruction Module}
\label{sec-method-rec}


During training, the reconstruction module separately takes two measurements $\ybm^r$ and $\ybm^m$ described in~\eqref{equ:pro-fwd} as inputs to produce two images $\xbmhat^r$ and $\xbmhat^m$ as outputs, respectively.  The measurements are first mapped to the image domain by applying the \markup{pseudoinverse} of their respective forward operators. We denote with ${(\Hbm^m)}^\dagger \ybm^m$ and ${(\Hbm^r)}^\dagger \ybm^r$ the resulting artifact-corrupted images in the image domain.
A CNN $\mathrm{h}_\thetabm$ with parameters $\thetabm\in\R^p$ is then trained to remove the artifacts from the corrupted images 
\begin{equation}
    \label{equ:recon}
    \xbmhat^m = \mathrm{h}_\thetabm \big({(\Hbm^m)}^\dagger \ybm^m\big) \ \text{and}\ \xbmhat^r = \mathrm{h}_\thetabm \big({(\Hbm^r)}^\dagger \ybm^r\big)\ .
\end{equation} 
Our network is a customized version of the residual CNN used in the prior work on deep image reconstruction~\cite{limEnhanced2017, yamanSelfsupervised2020, liuRARE2020}.

\markup{
Since the underlying true images $\xbm^m$ and $\xbm^r$ are unregistered, their reconstructed versions  $\xbmhat^m$ and $\xbmhat^r$ obtained from $\mathrm{h}_\thetabm$ are also unregistered. Therefore, it is suboptimal to construct a loss function to directly compare the pixel-wise difference between $\xbmhat^m$ and $\xbmhat^r$.}
It is thus necessary to use the registration module to mitigate their potential misalignment.
We define $\mathrm{T}(\xbmhat^r)$ and $\mathrm{T}(\xbmhat^m)$ as the images transformed according to the estimated deformation field (see details in Sec.~\ref{sec-method-reg}).
In our notation, $\mathrm{T}(\xbmhat^r)$ denotes a transformed variant of $\xbmhat^{r}$ relative to $\xbmhat^{m}$.


The loss function $\mathcal{L}_\mathrm{rec}$ of $\mathrm{h}_\thetabm$ has two components
\begin{equation}
    \label{equ:rec}
    \mathcal{L}_{\mathrm{rec}} = \mathcal{L}_{\mathrm{cross}} + \gamma\cdot \mathcal{L}_{\mathrm{self}}\ ,
\end{equation}
where the parameter $\gamma > 0$ controls the relative strength of each component.
The function $\mathcal{L}_\mathrm{cross}$ is the main component that penalizes the difference between the raw data and the transformed reconstructed image at a different motion state
\begin{equation}
    \begin{split}
    \mathcal{L}_\mathrm{cross} = \sum_{i = 1}^N \mathcal{L} \big(\ybm_{i}^r, \Hbm_{i}^r \Tsf(\xbmhat^{m}_i) \big) + \mathcal{L}\big(\ybm_{i}^m, \Hbm_{i}^m \Tsf(\xbmhat^{r}_i) \big)\ ,
    \end{split}
    \label{equ:rec_us}
\end{equation} 
where $\Hbm_{i}^m$ and $\Hbm_{i}^r$ are the forward operators used to map the registered images back to the measurement domain. Eq.~\eqref{equ:rec_us} maps pairs of measurements having the forms \eqref{equ:pro-fwd-r} and \eqref{equ:pro-fwd-m} by assuming that the deformations between them have been accounted for via the registration module. The function $\mathcal{L}_\mathrm{self}$ penalizes the discrepancy between the measurements estimated from a reconstructed image and the corresponding actual raw measurements 
\begin{equation}
    \label{equ:rec_self}
    \begin{split}
    \mathcal{L}_{\mathrm{self}} = \sum_{i = 1}^N\ \mathcal{L}\big(\ybm_{i}^r, \Hbm_{i}^r \xbmhat_{i}^r  \big) + \mathcal{L}\big(\ybm_{i}^m, \Hbm_{i}^m \xbmhat_{i}^m\big)\ .
    \end{split}
\end{equation}
Note that N2N/A2A can be seen as special cases of the proposed method where the potential deformations between the measurements are set to identity. 

\subsection{Registration Module}
\label{sec-method-reg}

Our registration module builds on self-supervised deep image registration discussed in Sec.~\ref{sec-background-registration}, which consists of a CNN $\mathrm{g}_\varphibm$, customized from U-net~\cite{ronnebergerUnet2015} with trainable parameters $\varphibm\in\R^q$, and a \emph{Spatial Transform Network (STN)}~\cite{jaderbergSpatial2015}.
As its order-sensitive input, the network accepts a pair of reconstructed images $(\xbmhat^{m}, \xbmhat^{r})$ estimated using $\mathrm{h}_\thetabm$ and registers them onto each other.
The network $\mathrm{g}_\varphibm$ uses two inputs in different orders to generate two motion fields
\begin{equation} 
    \label{equ:reg-g} 
    \begin{aligned}
        \phibmhat^{m \rightarrow r}=\mathrm{g}_\varphibm (\xbmhat^{m},\xbmhat^{r}) \ \text{and}\ \phibmhat^{r \rightarrow m}=\mathrm{g}_\varphibm\big(\xbmhat^{r},\xbmhat^{m})\ 
    \end{aligned}
\end{equation}
that characterize two coordinate mappings with opposite directions relative to each other.
For example, $\phibmhat^{m \rightarrow r}$ denotes a directional mapping from the coordinates of $\xbmhat^{m}$ to those of $\xbmhat^{r}$.
STN then transforms the coordinate of inputs based on the motion fields and obtains their registered variants 
\begin{equation}
    \label{equ:reg-stn} 
    \begin{aligned}
        \Tsf(\xbmhat^{m})=\xbmhat^{m} \circ \phibmhat^{m \rightarrow r} \ \text{and}\ \mathrm{T}(\xbmhat^{r})=\xbmhat^{r} \circ \phibmhat^{r \rightarrow m}\ .
    \end{aligned}
\end{equation}
The loss function  $\mathcal{L}_{\mathrm{reg}}$ for training $\gbm_\varphibm$ is specified as
\begin{equation}
    \mathcal{L}_{\mathrm{reg}} = \mathcal{L}_{\mathrm{similarity}} + \lambda\cdot\mathcal{L}_{\mathrm{smooth}}\ ,
\end{equation}
where $\mathcal{L}_{\mathrm{similarity}}$ enforces similarity between registered images and their references, $\mathcal{L}_{\mathrm{smooth}}$ enforces spatial smoothness in the motion field, and $\lambda > 0$ is a regularization parameter. The function $\mathcal{L}_{\mathrm{similarity}}$ is given by
\begin{equation}
    \label{equ:LLCC}
    \begin{aligned}
        \mathcal{L}_\mathrm{similarity} = -\sum_i\big( \mathrm{LCC}(\mathrm{T}(\xbmhat^{m}),\xbmhat^{r}_i) + \mathrm{LCC}(\mathrm{T}(\xbmhat^{r}),\xbmhat^{m}_i) \big).
    \end{aligned}
\end{equation}
where $\mathrm{LCC}$ denotes the local cross-correlation (LCC)~\cite{balakrishnanVoxelMorph2019}, which is known to be robust to intensity variations across different acquisitions~\cite{avantsSymmetric2008}.  While minimizing $L_\mathrm{similarity}$ enforces accurate alignment, it can also generate non-smooth registration fields that are not physically realistic~\cite{balakrishnanVoxelMorph2019}. Therefore, we include the function $\mathcal{L}_{\mathrm{smooth}}$ that imposes smoothness on the coordinate offsets $\hat{\vbm}=\phibmhat-\Ibm$
\begin{equation}
    \label{equ:smooth}
    \mathcal{L}_{\mathrm{smooth}} = \sum_i\Big(\norm{\Dbm\hat{\vbm}_i^{m \rightarrow r}}^2 + \norm{\Dbm\hat{\vbm}_i^{r \rightarrow m}}^2\Big)\ .
\end{equation}

\begin{figure}
    \centering
    \includegraphics[width=.48\textwidth]{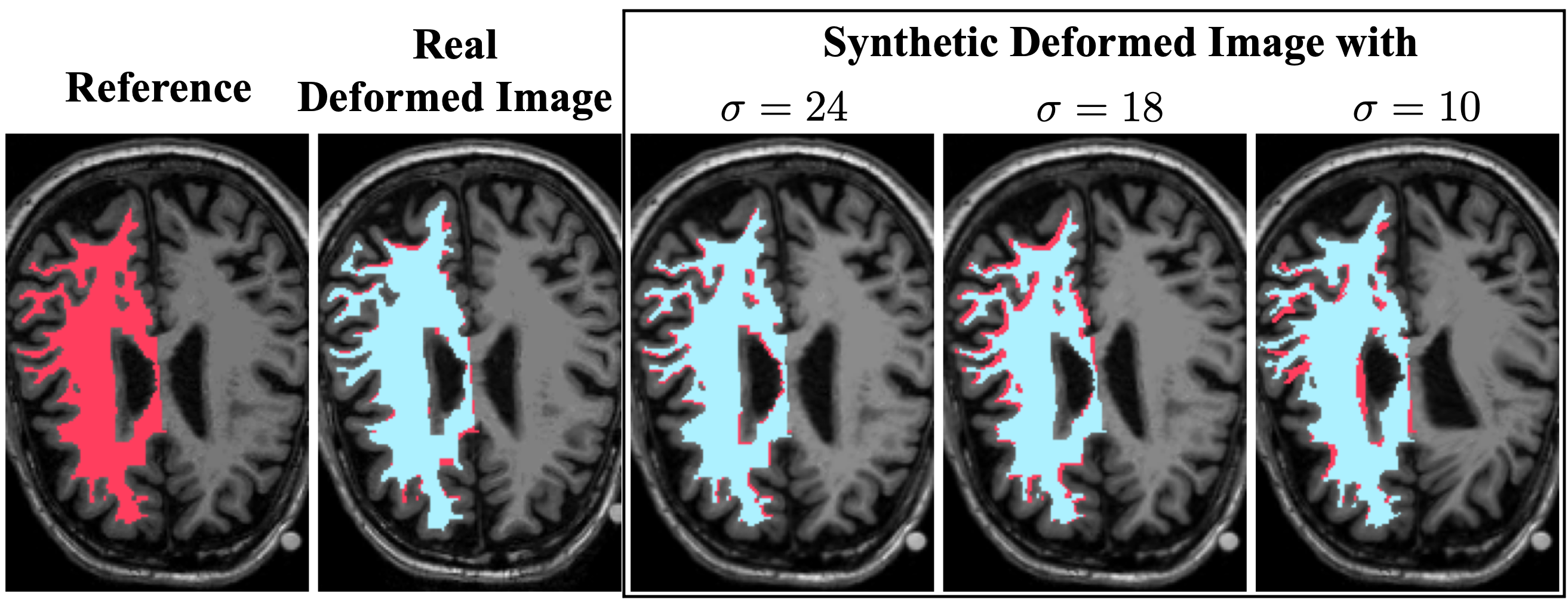}
    \caption{Visual illustration of deformations in the simulated experiments.
    The red regions are segmentations in the reference, while the blue regions are the corresponding segmentations in the deformed counterparts. 
    The synthetic deformations were generated by using the method in~\cite{sokootiNonrigid2017}, where $\sigma$ is inversely related to the deformation strengths.
    The \emph{in vivo} deformation is due to normal aging and disease. 
    }
    \label{fig:brain_setting}
\end{figure}

\begin{table*}
    \centering
    {
    \footnotesize
    \begin{threeparttable}
    \caption{\markup{Average PSNR and SSIM values \markup{obtained} over the test set.
    The table highlights that \markup{DeCoLearn} outperforms several well-known baseline methods at different acceleration factors and synthetic deformation magnitudes.
    }}
    \label{tb-simulated}
    \renewcommand\arraystretch{1}
    \setlength{\tabcolsep}{2pt}
    \begin{tabular}{ccccccccccccc} 
    \multicolumn{13}{c} {\textrm{Experiment of Simulated Measurement and Simulated Deformation}}\\
    \toprule
    \textit{Schemes}                                                & \multicolumn{6}{c}{PSNR}                                                                                       & \multicolumn{6}{c}{SSIM}                                                                                         \\ \cmidrule(rl){2-13}
    \textit{Synthetic Deformable with $\sigma=$}                    & \multicolumn{2}{c}{10}              & \multicolumn{2}{c}{18}              & \multicolumn{2}{c}{24}             & \multicolumn{2}{c}{10}              & \multicolumn{2}{c}{18}              & \multicolumn{2}{c}{24}               \\ 
    \cmidrule(rl){2-7}\cmidrule(rl){8-13}
    \textit{Acceleration rate}                                          & x3               & x4               & x3               & x4               & x3               & x4               & x3               & x4               & x3               & x4               & x3               & x4                \\ 
    \cmidrule(rl){2-3}\cmidrule(rl){4-5}\cmidrule(rl){6-7}\cmidrule(rl){8-9}\cmidrule(rl){10-11}\cmidrule(rl){12-13}
    Zero-Filled                                                & 28.20            & 26.03            & 28.19            & 26.02            & 28.28            & 26.05            & 0.772            & 0.717            & 0.772            & 0.715            & 0.774            & 0.716             \\
    Total Variation                                                 & 33.01            & 29.78            & 32.96            & 29.79            & 33.18            & 29.82            & 0.942            & 0.893            & 0.941            & 0.893            & 0.944            & 0.894             \\
    \markup{N2V~\cite{krullNoise2voidlearning2019}}  & \markup{28.19} & \markup{26.07}& \markup{28.19}& \markup{26.03}& \markup{28.35}& \markup{26.04}&\markup{0.774} & \markup{0.719}& \markup{0.774}&\markup{0.716} & \markup{0.778}&\markup{0.717} \\
    \markup{DIP~\cite{ulyanov2018deep}}           & \markup{32.64} &\markup{30.45} &\markup{32.87} &\markup{30.71} &\markup{33.05} &\markup{30.93} &\markup{0.913} &\markup{0.869} &\markup{0.915} &\markup{0.878} &\markup{0.915} &\markup{0.870} \\
    \markup{Self-Supervised}  &\markup{31.41} &\markup{29.58} &\markup{31.28} &\markup{28.92} & \markup{31.62} & \markup{29.74} &\markup{0.925} &\markup{0.922} &\markup{0.942} &\markup{0.910} &\markup{0.946} & \markup{0.908} \\
    \markup{SSDU~\cite{yaman2020self:mrm,yamanSelfsupervised2020}}  &\markup{32.98} &\markup{30.37} &\markup{32.92} &\markup{30.87} &\markup{33.13} &\markup{30.98} &\markup{0.956} &\markup{0.939} &\markup{0.954} &\markup{0.943} &\markup{0.959} &\markup{0.944} \\
    \markup{DeCoLearn}                                                           & \textbf{ 33.71 } & \textbf{ 31.60 } & \textbf{ 33.85 } & \textbf{ 31.67 } & \textbf{ 34.04 } & \textbf{ 31.72 } & \textbf{ 0.962 } & \textbf{ 0.945 } & \textbf{ 0.965 } & \textbf{ 0.947 } & \textbf{ 0.964 } & \textbf{ 0.949 }  \\ 
    \bottomrule
    \end{tabular}


    \end{threeparttable}
    }

\end{table*}

\begin{table*}
    \centering
    {
    \footnotesize
    \begin{threeparttable}
    \caption{\markup{Quantitative results of an ablation study showing influence of the registration module. The table shows that \markup{DeCoLearn} nearly matches the performance of the idealized \emph{A2A (Oracle)} method, which uses the true deformations.
    }}

    \label{tb-simulated-ablated}
    \renewcommand\arraystretch{1}
    \setlength{\tabcolsep}{2pt}
    \begin{tabular}{ccccccccccccc} 
    \multicolumn{13}{c} {\textrm{Experiment of Simulated Measurement and Simulated Deformation}}\\
    \toprule
    \textit{Schemes}                                                & \multicolumn{6}{c}{PSNR}                                                                                       & \multicolumn{6}{c}{SSIM}                                                                                         \\ \cmidrule(rl){2-13}
    \textit{Synthetic Deformable with $\sigma=$}                    & \multicolumn{2}{c}{10}              & \multicolumn{2}{c}{18}              & \multicolumn{2}{c}{24}             & \multicolumn{2}{c}{10}              & \multicolumn{2}{c}{18}              & \multicolumn{2}{c}{24}               \\ 
    \cmidrule(rl){2-7}\cmidrule(rl){8-13}
    \textit{Acceleration rate}                                          & x3               & x4               & x3               & x4               & x3               & x4               & x3               & x4               & x3               & x4               & x3               & x4                \\ 
    \cmidrule(rl){2-3}\cmidrule(rl){4-5}\cmidrule(rl){6-7}\cmidrule(rl){8-9}\cmidrule(rl){10-11}\cmidrule(rl){12-13}
    A2A (Unregistered)                                               & 30.19            & 29.07            & 31.96            & 30.37            & 32.83            & 30.89            & 0.921            & 0.903            & 0.942            & 0.926            & 0.954            & 0.935             \\
    A2A (Affine)                                                     & 30.42            & 29.14            & 32.50            & 30.67            & 33.42            & 31.20            & 0.922            & 0.900            & 0.950            & 0.932            & 0.959            & 0.940             \\
    A2A (SyN)                                                        & 32.70            & 30.31            & 32.71            & 30.35            & 32.85            & 30.39            & 0.952            & 0.929            & 0.957            & 0.932            & 0.956            & 0.933             \\
    A2A (VoxelMorph)                                                 & 32.44            & 30.26            & 33.06            & 30.67            & 33.16            & 31.03            & 0.950            & 0.928            & 0.958            & 0.936            & 0.957            & 0.938             \\ 
    \markup{DeCoLearn}                                                           & \textbf{ 33.71 } & \textbf{ 31.60 } & \textbf{ 33.85 } & \textbf{ 31.67 } & \textbf{ 34.04 } & \textbf{ 31.72 } & \textbf{ 0.962 } & \textbf{ 0.945 } & \textbf{ 0.965 } & \textbf{ 0.947 } & \textbf{ 0.964 } & \textbf{ 0.949 }  \\ 
    \cmidrule(lr){2-13}
    A2A (Oracle)$^\star$                                             & 34.17            & 31.89            & 34.20            & 31.91            & 34.29            & 31.93            & 0.965            & 0.948            & 0.965            & 0.948            & 0.966            & 0.949             \\
    \bottomrule
    \end{tabular}

    \begin{tablenotes}
        \item $^\star$: idealized algorithm, not available in practice.
    \end{tablenotes}

    \end{threeparttable}
    }

\end{table*}

\section{Experimental Validation}
\label{sec:exp}

We validate our method in the context of accelerated MRI. We consider three  settings: \emph{(a) 2D simulated measurements and simulated deformations}; \emph{(b) 2D simulated measurements and real unknown deformations}; and \emph{(c) 3D experimentally collected measurements and real unknown deformations}. 

\subsection{Setup}

\subsubsection{Baseline Methods}
\markup{
    We used several well-known image reconstruction methods for comparison
    \begin{enumerate}[label=(\alph*)]
            \item \emph{TV/CS}: The traditional total variation regularization method is summarized in eq.~\eqref{equ:optimization_tv}. On the experimentally collected free-breathing MRI data,  we replace the basic TV with the \emph{\markup{compressed} sensing (CS)} method from~\cite{eldenizConsistentlyAcquired2018}. Similarly to the well-known XD-GRASP method~\cite{fengXD2016}, CS exploits regularization along the motion dimension to significantly boost reconstruction performance.
            \item \emph{SSDU}/\emph{Self-Supervised}~\cite{yaman2020self:mrm}\footnote{We use the SSDU implementation at  \href{https://github.com/byaman14/SSDU}{github.com/byaman14/SSDU}.}: A recent self-supervised method that trains a \emph{deep unrolling network} by dividing each k-space MRI acquisition into two subsets and using them as training targets for each other. \emph{Self-Supervised} is a variant of SSDU that uses the same reconstruction CNN as DeCoLearn. Having both methods allows to separate the influence of the deep unrolling architecture from that of the training scheme on the SSDU performance.
            \item \emph{DIP/TDDIP}~\cite{yoo2021time}\footnote{We use the TDDIP implementation at \href{https://github.com/jaejun-yoo/TDDIP}{github.com/jaejun-yoo/TDDIP}.}: DIP is an image reconstruction method that uses an untrained CNN as a regularizer. We use an improved variant of DIP on our simulated data where two i.i.d.\ latent vectors are mapped to different measurements of the same subject. TDDIP is a recent extension of DIP that improves performance by taking into account the motion dimension in the image sequence. We use TDDIP on our experimentally-collected MRI data by sampling the latent inputs in the straight-line manifold due to the acyclic nature of the respiratory motion occurred in the dataset~\cite{yoo2021time}.
	\item \emph{Noise2Void (N2V)}~\cite{krullNoise2voidlearning2019}\footnote{We use the Noise2Void implementation at \href{https://github.com/juglab/n2v}{github.com/juglab/n2v}.}: An alternative to N2N that trains image restoration CNNs by mapping noisy pixels to their randomly-selected neighbors. Unlike N2N, N2V does not require paired data, but inherently assumes that artifacts are spatially unstructured---an assumption that does not hold for aliasing and streaking artifacts in MRI.
    \end{enumerate}
We also performed an ablation study to highlight the influence of the registration module within DeCoLearn. The ablated methods can be divided into three categories.}
\begin{itemize}
    \item {\bf Registration-free methods:}
    \begin{enumerate}[label=(\roman*)]
        \item \emph{A2A (Unregistered)}: The most basic variant of A2A, trained directly on unregistered measurements. It can be interpreted as the worst-case scenario for \markup{DeCoLearn} when no \markup{deformation}-compensation is performed during training.
    \end{enumerate}
    \item {\bf Pre-registration methods:} In this category, we explore \markup{the use of a fixed registration module that provides motion field estimates during the A2A training.}
    \begin{enumerate}[label=(\roman*)]
            \setcounter{enumi}{1}
            \item \emph{A2A (Affine)}: Uses \textsf{Affine} algorithms implemented in \emph{advanced normalization tools (ANTS)}~\cite{avants2009advanced}.
            \item \emph{A2A (SyN)}: Similar to \emph{A2A (Affine)}, but uses \emph{Symmetric Normalization (SyN)}~\cite{avantsSymmetric2008} algorithm instead.
            \item \emph{A2A (VoxelMorph)}: Uses a deep registration method from~\cite{balakrishnanVoxelMorph2019} pre-trained on artifact-corrupted images.
        \end{enumerate}
    \item {\bf Oracle-registration method:}
    \begin{enumerate}[label=(\roman*)]
            \setcounter{enumi}{4}
            \item \emph{A2A (Oracle)}: A2A (Oracle) is the idealized variant of \markup{DeCoLearn} using the registration model that provides perfect results. In our implementation, we synthesized the \emph{registered} data by applying different deformations and measurement operators on the same ground-truth image. Note that this method is \emph{not} applicable to the experimental data as the ground-truth is unavailable.
    \end{enumerate}
\end{itemize}

\begin{figure*}
    \begin{subfigure}{.565\textwidth}
        \centering
        \includegraphics[height=.305\textheight]{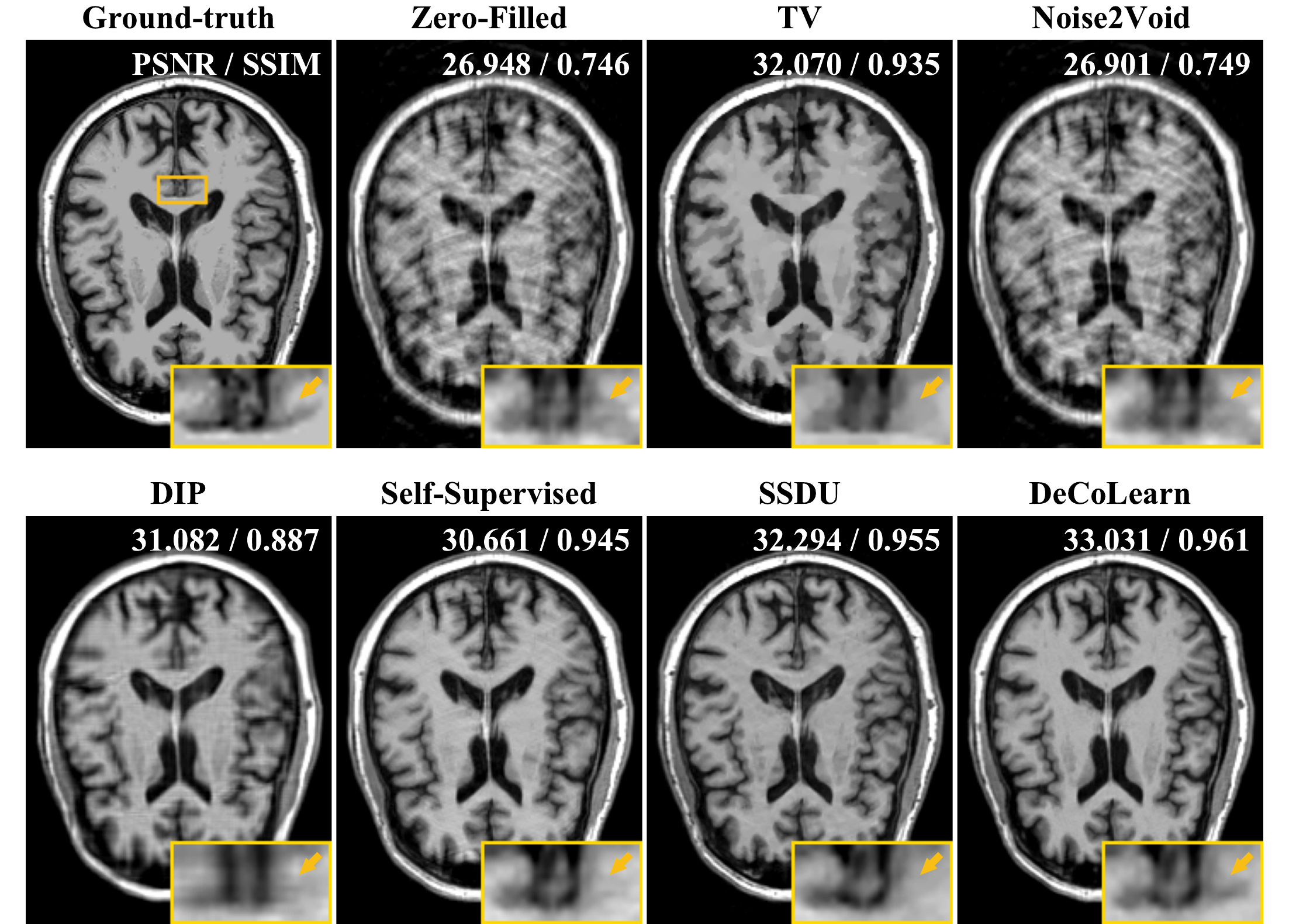}
        \caption{}
        \label{fig:brain-baseline}
    \end{subfigure}    
    \begin{subfigure}{.424\textwidth}
        \centering
        \includegraphics[height=.305\textheight]{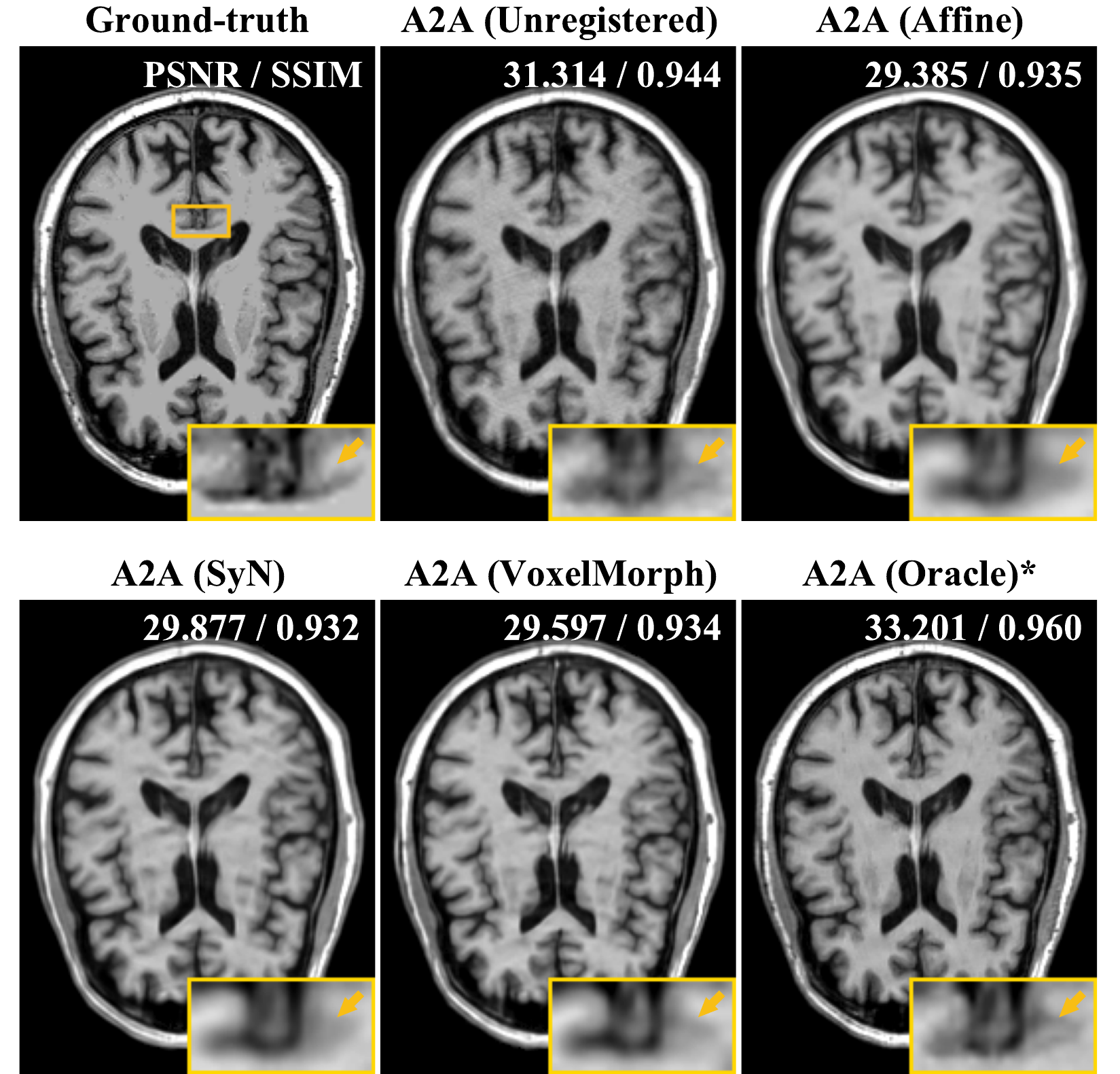}
        \caption{}
        \label{fig:brain-ablated}
    \end{subfigure} 
    \caption{
        \markup{Quantitative evaluation of DeCoLearn on simulated MRI measurements with \emph{in-vivo} deformations and 33\% sampling rate: \emph{(a)} comparison against other methods and \emph{(b)} results of an ablation study showing the influence of registration. The top-right corner of each image provides the PSNR and SSIM values with respect to the ground-truth. Yellow arrows in the highlight brain regions that were well reconstructed using DeCoLearn. Note that \emph{A2A (Oracle)} is an idealized algorithm that requires perfectly registered measurements that are unavailable in practice. This figure highlights that \markup{DeCoLearn} can achieve excellent quantitative and visual performance.}}
    \label{fig:brain}
\end{figure*}

\subsubsection{Evaluation Metrics}
In simulations, we implemented two \markup{widely-used} quantitative metrics, \emph{peak signal-to-noise ratio (PSNR)}, measured in dB and \emph{structural similarity index (SSIM)}, relative to the ground-truth images used to synthesize the measurements. 
\markup{Our evaluations on experimental data are qualitative due to the ground-truth being unavailable}.

\subsubsection{Implementation} 
We have experimented with several choices \markup{for the loss functions in eq.}~\eqref{equ:rec}.
The best empirical results were obtained when using the $\ell_1$ loss for the experimentally collected measurements, and the Huber function (or smooth-$\ell_1$ loss~\cite{renFaster2017}) for the simulated measurements. \markup{We set the learning rates of $\mathrm{Adam}_\mathrm{reg}$ and $\mathrm{Adam}_\mathrm{rec}$ to $0.0005$, and the mini-batch sizes to 4.} We performed all our experiments on a machine equipped with an Intel Xeon Gold 6130 Processor and an NVIDIA GeForce RTX 2080 Ti GPU.

\begin{table}
    \centering
    {
    \footnotesize
    \begin{threeparttable}
    \caption{\markup{Average PSNR and SSIM values obtained over the test set.
    Note how \markup{DeCoLearn} achieves better performance than all the methods at different acceleration factors.
    The deformations considered in this table are \emph{in vivo} due to normal aging and disease.
    }}
    \label{tb-simulated-real}

    \renewcommand\arraystretch{1}
    \setlength{\tabcolsep}{7pt}
    
    \begin{tabular}{ccccc} 
    \multicolumn{5}{c} {\textrm{Experiment of Simulated Measurement and Real Deformation}}\\
    \toprule
    \textit{Schemes}                  & \multicolumn{2}{c}{PSNR} & \multicolumn{2}{c}{SSIM}          \\ 
    \cmidrule(lr){2-5}
    \textit{Acceleration rate}            & x3    & x4                & x3    & x4                        \\ 
    \cmidrule(lr){2-3}\cmidrule(lr){4-5}
    Zero-Filled                  & 27.85 & 25.70             & 0.757 & 0.702                     \\
    Total Variation                   & 32.72 & 29.49             & 0.943 & 0.892                     \\
    \markup{N2V~\cite{krullNoise2voidlearning2019}}            &\markup{27.82} & \markup{25.69} & \markup{0.760} & \markup{0.703} \\
    \markup{DIP~\cite{ulyanov2018deep}}            & \markup{31.76} & \markup{30.70} & \markup{0.903} & \markup{0.876} \\
    \markup{Self-Supervised}            &\markup{31.16} & \markup{29.36} & \markup{0.950} & \markup{0.929} \\
    \markup{SSDU~\cite{yaman2020self:mrm,yamanSelfsupervised2020}}            &\markup{32.51} & \markup{30.18} & \markup{0.959} & \markup{0.945} \\
    \markup{DeCoLearn}                             & {\bf 33.23} & {\bf 31.19}             & {\bf 0.966} & {\bf 0.949}                     \\ 
    \bottomrule
    \end{tabular}

    \end{threeparttable}
    }

\end{table}

\begin{table}
    \centering
    {
    \footnotesize
    \begin{threeparttable}
    \caption{\markup{Quantitative results from an ablation study evaluating the influence of registration.
Note how \markup{DeCoLearn} achieves comparable performance to \emph{A2A (Oracle)}, which, unlike \markup{DeCoLearn}, relies on registration information obtained from the ground-truth. The deformations considered in this table are \emph{in vivo} due to normal aging and disease.
    }}
    \label{tb-simulated-real-ablated}

    \renewcommand\arraystretch{1}
    \setlength{\tabcolsep}{7pt}
    
    \begin{tabular}{ccccc} 
    \multicolumn{5}{c} {\textrm{Experiment of Simulated Measurement and Real Deformation}}\\
    \toprule
    \textit{Schemes}                  & \multicolumn{2}{c}{PSNR} & \multicolumn{2}{c}{SSIM}          \\ 
    \cmidrule(lr){2-5}
    \textit{Acceleration rate}            & x3    & x4                & x3    & x4                        \\ 
    \cmidrule(lr){2-3}\cmidrule(lr){4-5}
    A2A (Unregistered)                 & 31.94 & 30.05             & 0.953 & 0.932                     \\
    A2A (Affine)                       & 29.97 & 28.87             & 0.944 & 0.925                     \\
    A2A (SyN)                          & 30.66 & 28.88             & 0.947 & 0.921                     \\
    A2A (VoxelMorph)                   & 30.36 & 28.97             & 0.943 & 0/927                     \\
    \markup{DeCoLearn}                             & {\bf 33.23} & {\bf 31.19}             & {\bf 0.966} & {\bf 0.949}                     \\ 
    \cmidrule(lr){2-5}
    A2A (Oracle)$^\star$               & 33.85 & 31.52             & 0.966 & 0.949                     \\
    \bottomrule
    \end{tabular}

    \begin{tablenotes}
        \item $^\star$: unreachable without registered measurements.
    \end{tablenotes}

    \end{threeparttable}
    }

\end{table}

\begin{figure*}[t] 
    \centering
    \includegraphics[width=.975\textwidth]{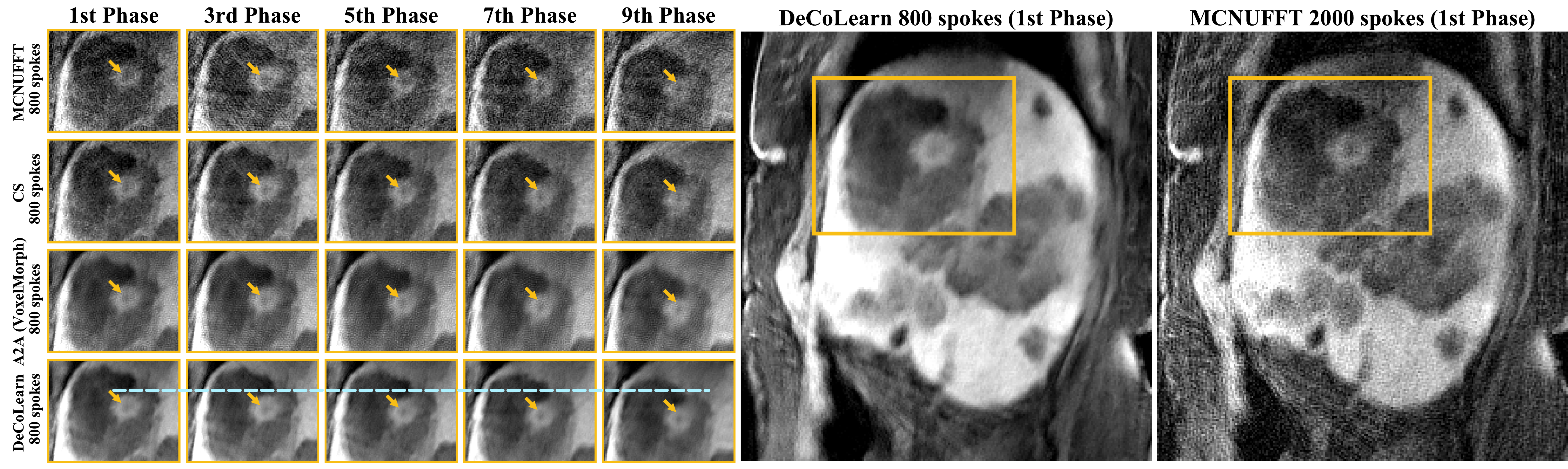}
    \caption{Illustration of \emph{in-vivo} respiratory deformation and several 3D reconstruction results from experimentally collected measurements corresponding to 800 spokes (about 2 minutes scan).
    The blue line provides a horizontal position reference of the tumor in the reconstruction result of \markup{DeCoLearn}, demonstrating nonrigid deformations between images across different respiratory phases.
    \markup{Yellow arrows indicate areas that were well preserved by DeCoLearn.}
    \markup{Note how DeCoLearn reconstructs higher quality images compared to both \emph{CS} and \emph{A2A (VoxelMorph)}.}}
    \label{fig:liver_setting}
\end{figure*} 

\subsection{Simulated Measurements and Deformations}
\label{sec-simulated}
\subsubsection{Dataset}\label{sec-simulated-data}
We used the T1-weighted MR brain acquisitions of 60 subjects obtained from the open dataset OASIS-3~\cite{lamontagneOASIS32019} as the raw \markup{ground-truth} for simulating measurements. These 60 subjects were split into 48, 6, and 6 for training, validation, and testing, respectively. For each subject, we extracted the middle 50 to 70 (depending on the shape of the brain) out of the 256 slices on the transverse plane, containing the most relevant regions of the brain. Each slice corresponds to $\xbm^r$ in \eqref{equ:pro-fwd-r}. We synthesized motion fields ($\phibm^{r \rightarrow m}$ in \eqref{equ:pro-fwd-m}) based on the method in~\cite{sokootiNonrigid2017} and used them to deform the ground-truth images, where the resulting images correspond to $\xbm^m$ in \eqref{equ:pro-fwd-m}. 
Three pre-defined parameters of the generation were the number of points randomly selected in the zero vector field $p=2000$, the range of random values assigned to those points $\delta=[-10, 10]$, and the standard deviations of the smoothing Gaussian kernel for the vector field $\sigma\in\{10,18, 24\}$. Thus, $\sigma$ is inversely related to the strength of deformation in the image.
Fig.~\ref{fig:brain_setting} shows visual examples of the deformed images generated by synthetic registration fields with different values of $\sigma$.
In order to obtain corrupted measurement pairs, we simulated a single-coil MRI setting with a \markup{Cartesian sampling pattern that sub-samples and fully-samples along $ky$ and $kx$ dimension in the k-space, respectively}.
We set the sampling rate to 25\% and 33\% (corresponding to $4\times$ and $3\times$ acceleration) of the full sampling rate for the complete k-space data and added measurement noise corresponding to an input SNR of 40dB.

\subsubsection{Results}
\markup{Table~\ref{tb-simulated} summarizes quantitative results of all the evaluated methods.
Note that the improvement of \emph{SSDU} over \emph{Self-Supervised} is due to the \emph{deep unrolling} architecture, that, in principle, can also be adopted in DeCoLearn to further improve its performance.
Table~\ref{tb-simulated} shows that \markup{DeCoLearn} achieves the highest PSNR and SSIM values compared to other methods over all considered configurations of subsampling and deformation strengths.
Table~\ref{tb-simulated-ablated} shows the quantitative results of the ablation study evaluating the influence of the deep registration module. The results suggest that pre-registering images before training leads to sub-optimal performance, while DeCoLearn nearly matches the performance of the idealized \emph{A2A (Oracle)} that uses the ground-truth deformations.}

\subsection{Simulated Measurements and Real Deformations}
\label{sec-simulated-real}
\subsubsection{Dataset}
We consider a data acquisition scheme that is similar to that described in Sec.~\ref{sec-simulated}, but differs in the approach to deform the \markup{ground-truth}.
Specifically, we used the second MR acquisitions of the 60 subjects from the OASIS-3~\cite{lamontagneOASIS32019} dataset as the deformed images.
The intervals between the two MR sessions of each subject range from one to ten years. Note that the deformations occurring in two different \emph{in vivo} MR images of the same subject are due to normal aging and the potential effects of the Alzheimer disease. Fig.~\ref{fig:brain_setting} visually illustrates the corresponding deformation.

\subsubsection{Results}
\markup{
    Fig.~\ref{fig:brain-baseline} summarizes the results from all the evaluated methods on this dataset. One can observe a significant reduction in imaging artifacts due to \emph{TV} compared to the \emph{Zero-Filled} reconstruction. However, \emph{TV} also leads to a loss of detail due to the well-known ``staircase effect''. The poor relative performance of \emph{N2V} implies that its effectiveness in image denoising does not translate well to the removal of structured aliasing artifacts. The yellow arrows in the magnified regions of Fig.~\ref{fig:brain-baseline} highlight brain tissue that was clearly reconstructed using only DeCoLearn. 

    Fig.~\ref{fig:brain-ablated} provides results from the ablation study. Pre-registration methods, such as \emph{A2A (VoxelMorph)}, lead to a significant improvements over the registration-free methods by using pre-registered artifact-contaminated images, but they still suffer from smoothing in the region indicated by yellow arrows. \markup{DeCoLearn} achieves better performance compared to all of these ablated methods in terms of sharpness, contrast, and artifact removal, due to its ability to correct for deformations during training. Note that although the measurements were simulated in this experiment for quantitative evaluation, the deformations in the data are \emph{in vivo}.
}

\begin{figure*}[t]
\centering 
\includegraphics[width=.79\textwidth]{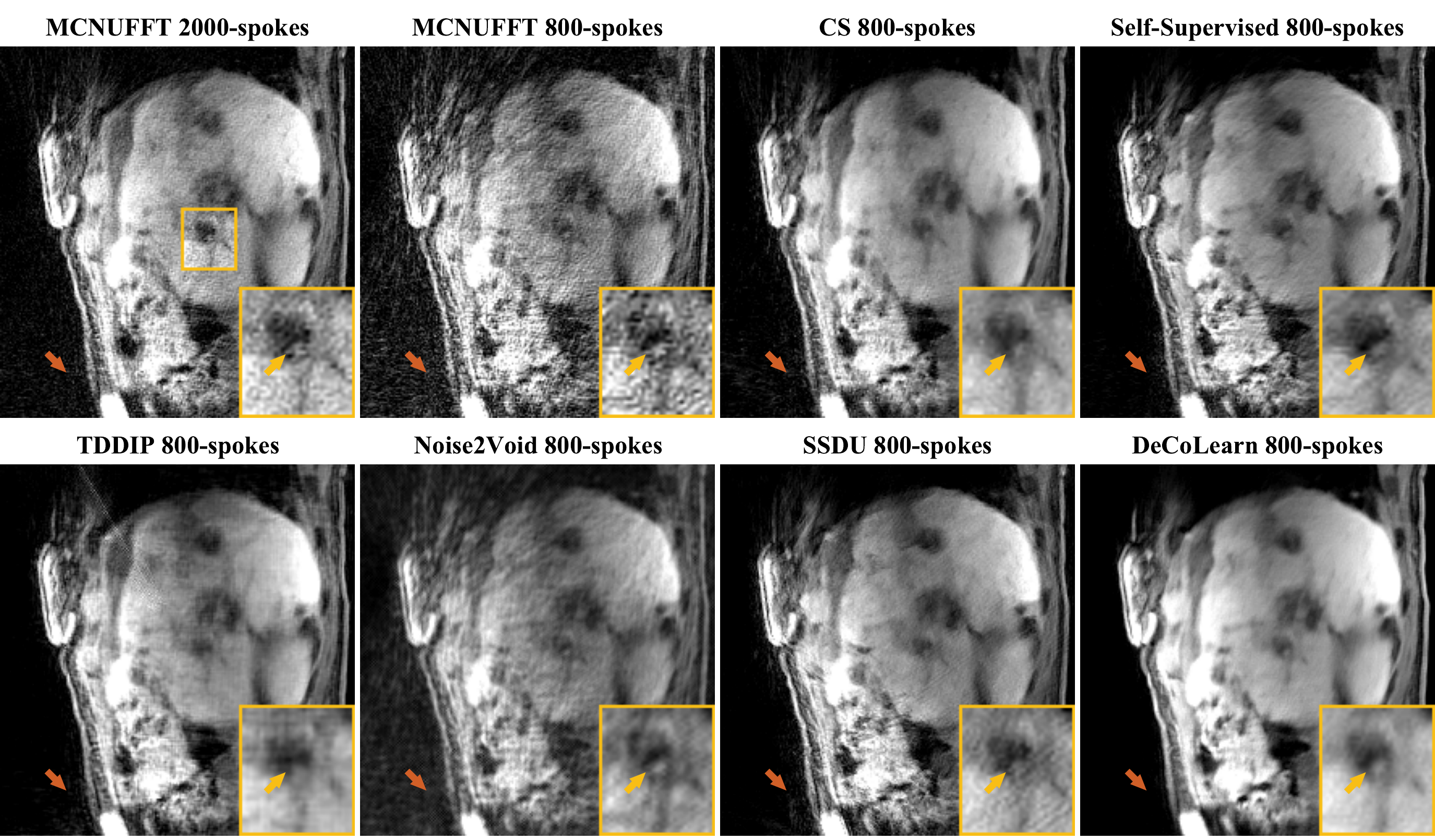}
\caption{\markup{Comparison of several reconstruction methods on experimentally collected data corresponding to 800 radial spokes (scans of about 2 minutes). N2V, SSDU, and Self-Supervised are all trained by using the available 800 spokes at each motion state. 
CS and TDDIP take advantage of the correlations in the respiratory motion dimension by imposing an additional regularizer and encoding the motion trajectory into input latent variables, respectively. DeCoLearn improves over A2A training by correcting for deformations in different motion states. The visually important differences are highlighted using arrows. Note how compared to other methods, DeCoLearn recovers sharper images (see yellow arrows in magnified regions) and reduces artifacts (see orange arrows in the background). 
}}
 \label{fig:liver_recon_general}
\end{figure*}

\begin{figure*}[t] 
    \centering 
    \includegraphics[width=\textwidth]{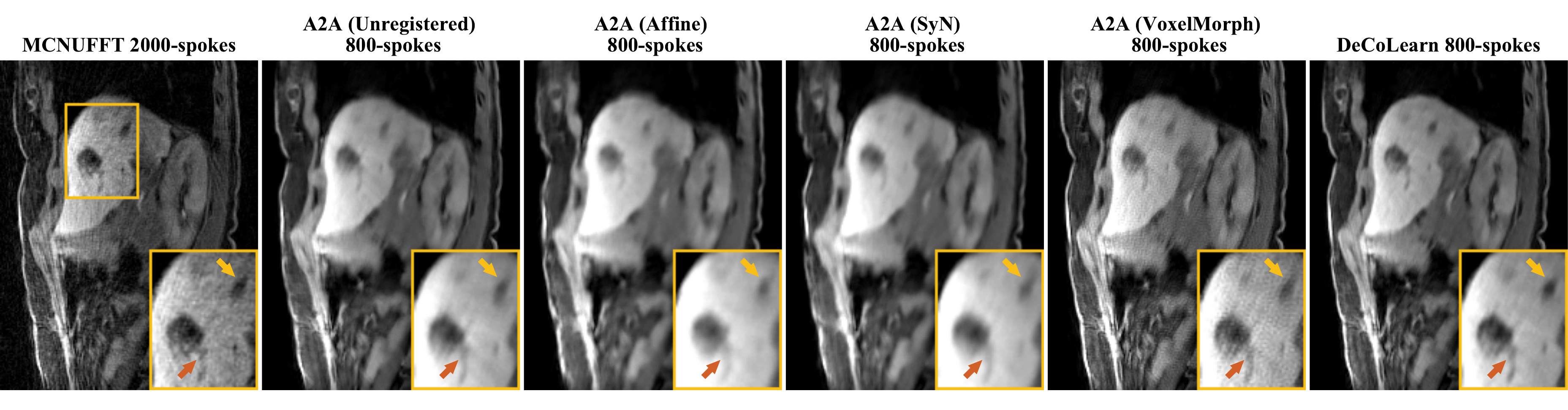}
    \caption{
        \markup{Illustration of the results from the ablation study of DeCoLearn on experimentally-collected data corresponding to 800 radial spokes (scans of about 2 minutes). \emph{A2A (Unregistered)} is directly trained on unregistered 3D measurement pairs, while \emph{A2A (SyN)} and \emph{A2A (VoxelMorph)} train CNNs on pre-registered but artifact-corrupted images. \emph{MCNUFFT 2000-spokes} requires data corresponding to 2000 radial spokes (scans of about 5 minutes). The visual differences are highlighted using arrows in magnified regions. Note how DeCoLearn outperforms its ablated variants by jointly performing 3D image reconstruction and registration.}
    }
    \label{fig:liver_recon_aba}
\end{figure*}

\subsection{Real Measurements and Real Deformations}
\label{real-experiment}
\subsubsection{Dataset} 
All acquisition processes were performed on a 3T PET/MRI scanner (Biograph mMR; Siemens Healthcare, Erlangen, Germany).
We collected the data by using the CAPTURE method, a T1-weighted stack-of-stars 3D spoiled gradient-echo sequence with fat suppression that has consistently acquired projections for respiratory motion detection~\cite{eldenizConsistentlyAcquired2018}.
The acquisition parameters were as follows: TE/TR $=$ 1.69ms/3.54ms, \markup{FOV $=$ 360 $\times$ 360 $\times$ 288 - 360 $\times$ 360 $\times$ 360 $\text{mm}^3$, in-plane resolution=1.125 $\times$ 1.125 $\times$ 3 $\text{mm}^3$}, partial Fourier factor $=$ 6/8, number of radial spokes $=$ 2000, slice resolution $=$ 50\%, slice per slab $N_z=\{$96, 112, 120$\}$ so as to cover the torso with \markup{an interpolated} slice thickness of 3mm, total acquisition time was about 5 minutes (slightly longer for larger subjects). 
We discarded the first ten spokes during reconstruction to ensure the acquired signal reached a steady state. 
Our free-breathing MRI data were subsequently binned into $N_p=10$ respiratory phases, and thus each phase was reconstructed with $N_s=199$ spokes.
The dimension of raw measurement for each subject was $N_z\times N_c\times N_p\times N_s\times N_l$ with $N_c=\{5,6\}$ being the number of coils and $N_l$ being the length of radial spokes.
The coil sensitivity maps were estimated from the central radial k-space spokes of each slice and were assumed to be known during experiments.
Apodization was applied by using a Hamming window that covered the central k-space in order to avoid Gibbs ringing. 
We used inverse Multi-Coil Non-Uniform Fast Fourier Transform (MCNUFFT)~\cite{muckley:20:tah} to map those measurements from k-space to the image domain, yielding 4D images $N_x\times N_y\times N_p\times N_z$ for each subject where $N_x\times N_y$ is the image domain matrix size.

\begin{figure*}
\centering
\includegraphics[width=\textwidth]{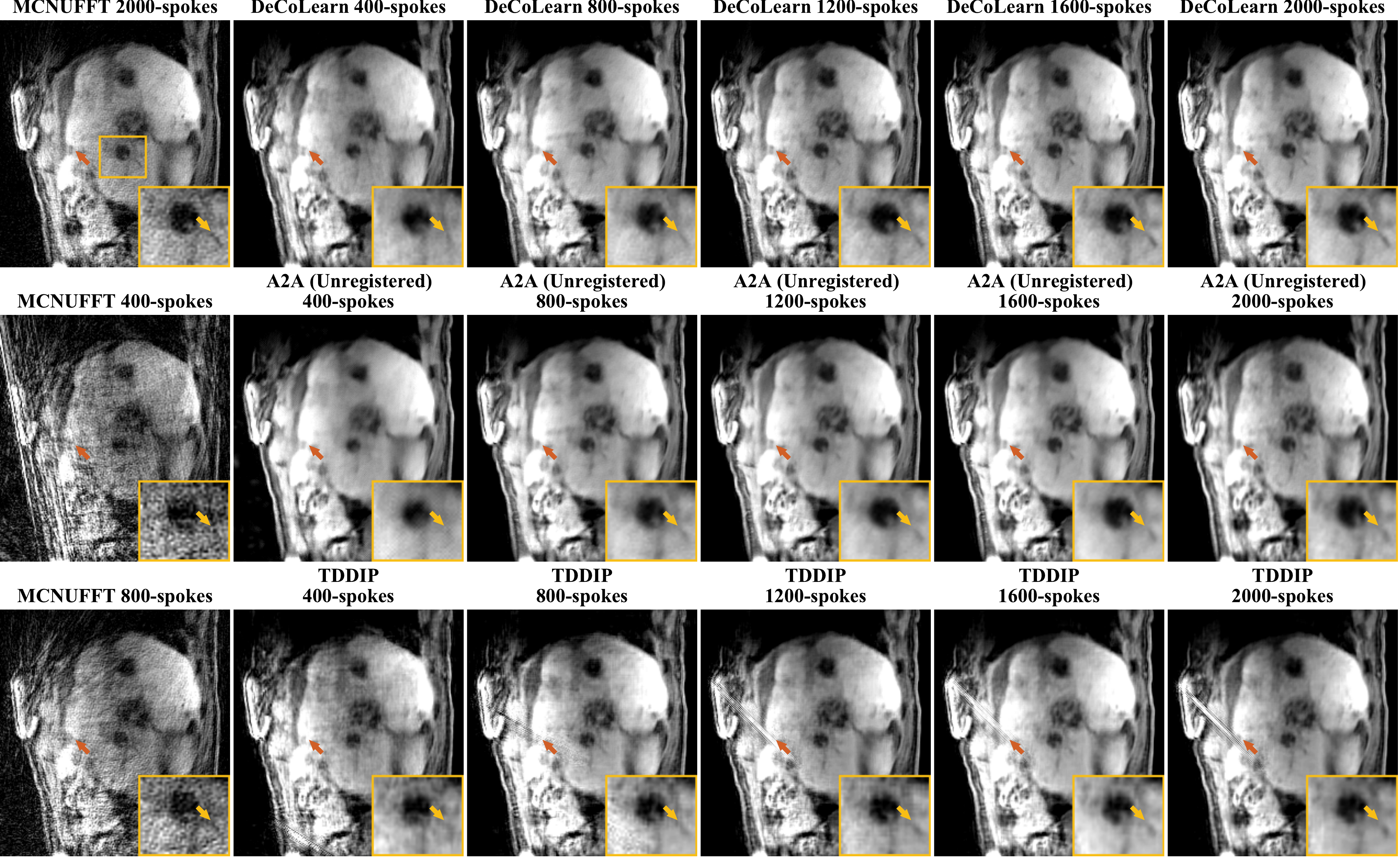}
\caption{Illustration of reconstruction results of \markup{DeCoLearn, \emph{A2A (Unregistered)}, and \emph{TDDIP}} from experimentally collected measurements using 400, 800, 1200, 1600, and 2000 spokes, corresponding to 1-, 2-, 3-, 4-, and 5- minute scans, respectively. \markup{\emph{A2A (Unregistered)} trains CNNs on unregistered measurements. \emph{TDDIP} is a variant of DIP that improves performance jointly reconstructing images of 10 respiratory phases. We highlighted visual differences by using arrows. Note how DeCoLearn reconstructs sharper edges (see liver tissues highlighted by yellow arrows in the magnified region) and better reduces artifacts (see image backgrounds highlighted by orange arrows).} This figure shows that \markup{DeCoLearn} can improve over these two methods at different acquisitions durations by integrating a deep image registration module.}
    \label{fig:liver_recon_diff_nLines}
\end{figure*}  

Upon the approval of our Institutional Review Board, multichannel liver data from ten healthy volunteers and six cancer patients were used in this paper, where eight healthy subjects were used for training, one healthy subject for validation, and the rest for testing. Raw measurements of each subject were first reformatted into $N_z$ measurements, yielding $8N_z$ samples for training and $N_z$ for validation.
We then trained \markup{DeCoLearn} on measurement pairs such that each pair contained the five odd \markup{respiratory} phases and the five even \markup{respiratory} phases of the same training sample.
Fig.~\ref{fig:liver_setting} shows examples of MCNUFFT images obtained from a training sample, demonstrating that \markup{DeCoLearn} was trained on unregistered measurement pairs corresponding to images with nonrigid respiratory deformations.
We used MCNUFFT images from the full acquisition duration (5 minutes) as the reference for qualitative evaluations.
We conducted the experiments for various acquisition durations of 1, 2, 3, 4, and 5 minutes, corresponding to 400, 800, 1200, 1600, and 2000 radial spokes in k-space, respectively. \markup{The golden-angle acquisition scheme ensures approximately \emph{uniform} coverage of k-space for any arbitrary number of consecutive spokes~\cite{feng2014golden}.}

\markup{The original implementation of SSDU is based on the \emph{fast Fourier transform (FFT)},  which is not suitable to the non-uniform sampling pattern used in our data. Therefore, we re-implemented SSDU by using a publicly available non-uniform FFT operator~\cite{muckley:20:tah} and the \emph{unrolled regularization by denoising} architecture~\cite{liu2021sgd}. Though \emph{Self-Supervised} relies on the same 3D network as DeCoLearn, due to memory constraints, \emph{SSDU} is implemented as a 2D architecture that processes each individual phase separately. Note that the original implementation of \emph{SSDU} is also based on a 2D architecture.}

\subsubsection{Results}
\markup{
Fig.~\ref{fig:liver_recon_general} shows reconstruction results of all the methods on 800 radial spokes (corresponding to about 2 minute acquisitions). The \emph{MCNUFFT} image suffers from strong streaking artifacts. Note how even \emph{MCNUFFT 2000-spokes}, which corresponds to about 5 minute acquisitions, leads to imaging artifacts. All other methods yield significant improvements over \emph{MCNUFFT}. While the result of \emph{CS} (which is similar to the well-known XD-GRASP method) shows a considerable reduction in the streaking artifacts, it also contains a noticeable amount of detail loss. \emph{N2V} reduces the noise-like artifacts, but still preserves the structured streaking artifacts. The results of \emph{SSDU} and \emph{Self-Supervised} show the benefit of N2N-type of training over that of N2V for image reconstruction. Overall, DeCoLearn achieves the best qualitative performance. As highlighted in Fig.~\ref{fig:liver_recon_general} using arrows, DeCoLearn reconstructs sharper edges (see yellow arrows) and reduces background imaging artifacts (see orange arrows).
}

\markup{Fig.~\ref{fig:liver_recon_aba} illustrates the results of the ablation experiments on the real data with 800 radial spokes.}
\emph{A2A (Unregistered)} leads to a reasonable result even without registration in training, but it also contains a noticeable amount of blur, especially along the edges.
\emph{A2A (Affine)} and \emph{A2A (SyN)} also suffer from smoothing in the region of interest even with the registration algorithms integrated to pre-align the samples.
Note the reduction in blur in \emph{A2A (VoxelMorph)} relative to the registration-free methods. However, a closer inspection indicates that \markup{the result of \emph{A2A (VoxelMorph)}} still suffers from artifacts, \markup{such as the noise-like artifacts around the spot highlighted by yellow and orange arrows}. Fig.~\ref{fig:liver_recon_aba} depicts that \markup{DeCoLearn} leads to improvements over several baseline methods, especially compared with \emph{MCNUFFT 2000 spokes} with a longer acquisitions time (5 minutes).
Fig.~\ref{fig:liver_setting} also provides visual comparisons between \markup{DeCoLearn}, \emph{CS} and \emph{A2A (VoxelMorph)}. \markup{Fig.~\ref{fig:liver_setting} shows that \markup{DeCoLearn} performs better across different respiratory phases, especially considering its ability to remove artifacts around the spot highlighted by yellow arrows.} Note that both the measurements and the deformations in these results are from experimentally collected data, demonstrating the applicability of \markup{DeCoLearn} in motion-\markup{resolved} MRI.

Fig.~\ref{fig:liver_recon_diff_nLines} illustrates comparisons between \emph{A2A (Unregistered)}\markup{, \emph{TDDIP} and DeCoLearn} for various acquisition durations.
\markup{We annotated visual differences using yellow and red arrows.}
While \emph{A2A (Unregistered)} trains CNNs directly on unregistered measurement pairs, \markup{\markup{DeCoLearn} reconstructs sharper boundaries highlighted by yellow arrows due to its ability to take into account the deformation field during training.
These results indicates the excellent performance of DeCoLearn across different acquisition durations.}

\section{\markup{Discussion and Conclusion}}
\subsection{\markup{Benefits of DeCoLearn}} 

\markup{DeCoLearn enables learning using information from multiple measurements of the same object undergoing nonrigid deformation.
Unlike N2N/A2A, DeCoLearn relaxes the requirement on having registered measurements, making it more applicable in practice.
DeCoLearn is fully complementary to existing self-supervised methods that use a single measurement, such as SSDU~\cite{yaman2020self:mrm,yamanSelfsupervised2020} and N2V~\cite{krullNoise2voidlearning2019}. One can simply integrate DeCoLearn with these self-supervised schemes by imposing an additional self-supervision term.
Note also that DeCoLearn is compatible with any deep unrolling architecture.
}

\subsection{\markup{Potential Extension and Future Works}}

\subsubsection{\markup{Extension to Contrast-Variant Measurements}}
\markup{The current implementation of DeCoLearn can only compensate image deformations over different acquisitions of the same object. In some dynamic imaging scenarios, such as the dynamic contrast enhanced (DCE) imaging~\cite{gordon2014dynamic}, different measurements acquired from the same object might also correspond to distinct image contrasts. DeCoLearn is not yet suitable for such imaging problems. Extension of DeCoLearn to this scenario would be an interesting direction of future research.}

\subsubsection{\markup{Extension to Sequential Image Reconstruction}}
\markup{The reconstruction of a sequence of images from the measurements of a dynamic object has many applications in medical imaging (e.g., cine dynamic imaging). The key concept behind dynamic imaging is to leverage the redundancies in the data across the motion dimension (see our discussion of \emph{MoCo} reconstruction). Our experimental validation on free-breathing MRI has shown that DeCoLearn can be used to learn the redundancies over the respiratory dimension. However, DeCoLearn does not explicitly properties specific to the motion dimension. Future work can address this by extending DeCoLearn to include an explicit motion regularization.}

\subsection{Conclusion}
\label{sec-conclusion}
We proposed a new method for addressing an important issue in the context of training of deep neural networks for medical image reconstruction. Our proposed \markup{DeCoLearn} method  extends the influential Noise2Noise approach by working directly in the measurement domain and compensating for object motion in the data. We validated our method using simulated and experimentally collected MRI data. Our results demonstrated that \markup{DeCoLearn} significantly \markup{improves} image quality compared to several baseline methods. Though our experiments focused on MRI, the DeCoLearn method has the potential to be adopted in other imaging modalities as well, such as computerized tomography~\cite{lei4dct2019} and optical diffraction tomography~\cite{kamilovOptical2016}. \markup{In such imaging scenarios}, it is often impossible to obtain fully-sampled measurements, but only several distinct views of the object \markup{where it is possible that these views are not registered onto each other}.


\begin{thebibliography}{10}
    
    \bibitem{lustigSparse2007}
    M.~Lustig, D.~Donoho, and J.~M. Pauly, ``Sparse {{MRI}}: {{The}} application of
      compressed sensing for rapid {{MR}} imaging,'' \emph{Magn. Reson. Med.},
      vol.~58, no.~6, pp. 1182--1195, 2007.
    
    \bibitem{danielyanBM3D2011}
    A.~Danielyan, V.~Katkovnik, and K.~Egiazarian, ``{{BM3D}} frames and
      variational image deblurring,'' \emph{IEEE Trans. Image Process.}, vol.~21,
      no.~4, pp. 1715--1728, 2011.
    
    \bibitem{eladImage2006}
    M.~Elad and M.~Aharon, ``Image denoising via sparse and redundant
      representations over learned dictionaries,'' \emph{IEEE Trans. Image
      Process.}, vol.~15, no.~12, pp. 3736--3745, 2006.
    
    \bibitem{huFast2011}
    Y.~Hu, S.~G. Lingala, and M.~Jacob, ``A fast majorize\textendash minimize
      algorithm for the recovery of sparse and low-rank matrices,'' \emph{IEEE
      Trans. Image Process.}, vol.~21, no.~2, pp. 742--753, 2011.
    
    \bibitem{rudinNonlinear1992}
    L.~I. Rudin, S.~Osher, and E.~Fatemi, ``Nonlinear total variation based noise
      removal algorithms,'' \emph{Physica D}, vol.~60, no. 1-4, pp. 259--268, 1992.
    
    \bibitem{knollDeeplearning2020}
    F.~Knoll, K.~Hammernik, C.~Zhang, S.~Moeller, T.~Pock, D.~K. Sodickson, and
      M.~Akcakaya, ``Deep-learning methods for parallel magnetic resonance imaging
      reconstruction: {{A}} survey of the current approaches, trends, and issues,''
      \emph{IEEE Signal Process. Mag.}, vol.~37, no.~1, pp. 128--140, 2020.
    
    \bibitem{lucasUsing2018}
    A.~Lucas, M.~Iliadis, R.~Molina, and A.~K. Katsaggelos, ``Using deep neural
      networks for inverse problems in imaging: Beyond analytical methods,''
      \emph{IEEE Signal Process. Mag.}, vol.~35, no.~1, pp. 20--36, 2018.
    
    \bibitem{mccannConvolutional2017}
    M.~T. McCann, K.~H. Jin, and M.~Unser, ``Convolutional neural networks for
      inverse problems in imaging: {{A}} review,'' \emph{IEEE Signal Process.
      Mag.}, vol.~34, no.~6, pp. 85--95, 2017.
    
    \bibitem{ongieDeep2020}
    G.~Ongie, A.~Jalal, C.~A. Metzler, R.~G. Baraniuk, A.~G. Dimakis, and
      R.~Willett, ``Deep learning techniques for inverse problems in imaging,''
      \emph{IEEE J. Sel. Areas Inf. Theory}, vol.~1, no.~1, pp. 39--56, 2020.
    
    \bibitem{wangDeep2020}
    G.~Wang, J.~C. Ye, and B.~De~Man, ``Deep learning for tomographic image
      reconstruction,'' \emph{Nat. Mach. Intell.}, vol.~2, no.~12, pp. 737--748,
      Dec. 2020.
    
    \bibitem{akccakaya2021unsupervised}
    M.~Ak{\c{c}}akaya, B.~Yaman, H.~Chung, and J.~C. Ye, ``Unsupervised deep
      learning methods for biological image reconstruction,''
      \emph{arXiv:2105.08040}, 2021.
    
    \bibitem{lehtinenNoise2Noise2018}
    J.~Lehtinen, J.~Munkberg, J.~Hasselgren, S.~Laine, T.~Karras, M.~Aittala, and
      T.~Aila, ``{{Noise2Noise}}: {{Learning}} image restoration without clean
      data,'' in \emph{Proc. Int. Conf. Machine Learning}, 2018.
    
    \bibitem{liuRARE2020}
    J.~Liu, Y.~Sun, C.~Eldeniz, W.~Gan, H.~An, and U.~S. Kamilov, ``{RARE}: Image
      reconstruction using deep priors learned without ground truth,'' \emph{IEEE
      J. Sel. Top. Signal Process.}, 2020.
    
    \bibitem{krullNoise2voidlearning2019}
    A.~Krull, T.-O. Buchholz, and F.~Jug, ``{Noise2Void}-learning denoising from
      single noisy images,'' in \emph{Proc. IEEE Conf. Computer Vision and Pattern
      Recognition}, 2019, pp. 2129--2137.
    
    \bibitem{yamanSelfsupervised2020}
    B.~Yaman, S.~A.~H. Hosseini, S.~Moeller, J.~Ellermann, K.~U{\u g}urbil, and
      M.~Ak{\c c}akaya, ``Self-supervised physics-based deep learning {{MRI}}
      reconstruction without fully-sampled data,'' in \emph{Proc. Int. Symp.
      Biomedical Imaging}, 2020, pp. 921--925.
    
    \bibitem{laineHighquality2019}
    S.~Laine, T.~Karras, J.~Lehtinen, and T.~Aila, ``High-quality self-supervised
      deep image denoising,'' \emph{Advances in Neural Information Processing
      Systems}, vol.~32, pp. 6970--6980, 2019.
    
    \bibitem{ganPhase2Phaseinpress}
    C.~Eldeniz, W.~Gan, S.~Chen, T.~J. Fraum, D.~R. Ludwig, Y.~Yan, J.~Liu,
      T.~Vahle, U.~B. Krishnamurthy, U.~S. Kamilov, and H.~An, ``{{Phase2Phase}}:
      Respiratory motion-resolved reconstruction of free-breathing {MRI} using deep
      learning without a ground truth for improved liver imaging,'' \emph{Invest.
      Radiol.}, 2021.
    
    \bibitem{torop2020deep}
    M.~Torop, S.~V. Kothapalli, Y.~Sun, J.~Liu, S.~Kahali, D.~A. Yablonskiy, and
      U.~S. Kamilov, ``Deep learning using a biophysical model for robust and
      accelerated reconstruction of quantitative, artifact-free and denoised
      images,'' \emph{Magn. Reson. Med.}, vol.~84, no.~6, pp. 2932--2942, 2020.
    
    \bibitem{ehret2019model}
    T.~Ehret, A.~Davy, J.~Morel, G.~Facciolo, and P.~Arias, ``Model-blind video
      denoising via frame-to-frame training,'' in \emph{Proc. IEEE Conf. Computer
      Vision and Pattern Recognition}, 2019, pp. 11\,369--11\,378.
    
    \bibitem{yu2020joint}
    S.~Yu, B.~Park, J.~Park, and J.~Jeong, ``Joint learning of blind video
      denoising and optical flow estimation,'' in \emph{Proc. IEEE Conf. Computer
      Vision and Pattern Recognition Workshops}, 2020, pp. 500--501.
    
    \bibitem{jiang2020weakly}
    Z.~Jiang, Z.~Huang, B.~Qiu, X.~Meng, Y.~You, X.~Liu, M.~Geng, G.~Liu, C.~Zhou,
      K.~Yang \emph{et~al.}, ``Weakly supervised deep learning-based optical
      coherence tomography angiography,'' \emph{IEEE Trans. Med. Imaging}, vol.~40,
      no.~2, pp. 688--698, 2020.
    
    \bibitem{buchholzCryoCARE2019}
    T.-O. Buchholz, M.~Jordan, G.~Pigino, and F.~Jug,
      ``{Cryo-{{CARE}}: Content-aware image restoration for
      cryo-transmission electron microscopy data},'' in
      \emph{{Proc. Int. Symp. Biomedical Imaging}}, Apr.
      2019, pp. 502--506.
    
    \bibitem{hendriksenNoise2Inverse2020}
    A.~A. Hendriksen, D.~M. Pelt, and K.~J. Batenburg,
      ``{{{Noise2Inverse}}: Self-supervised deep
      convolutional denoising for tomography},'' \emph{{IEEE
      Trans. Comput. Imaging}}, vol.~6, pp. 1320--1335, 2020.
    
    \bibitem{xu2021deformed2self}
    J.~Xu and E.~Adalsteinsson, ``{Deformed2Self}: Self-supervised denoising for
      dynamic medical imaging,'' \emph{arXiv:2106.12175}, 2021.
    
    \bibitem{fuDeep2020}
    Y.~Fu, Y.~Lei, T.~Wang, W.~J. Curran, T.~Liu, and X.~Yang, ``Deep learning in
      medical image registration: A review,'' \emph{Phys. Med. Biol.}, vol.~65,
      no.~20, p. 20TR01, 2020.
    
    \bibitem{devosDeep2019}
    B.~D. {de Vos}, F.~F. Berendsen, M.~A. Viergever, H.~Sokooti, M.~Staring, and
      I.~I{\v s}gum, ``A deep learning framework for unsupervised affine and
      deformable image registration,'' \emph{Med. Image Anal.}, vol.~52, pp.
      128--143, 2019.
    
    \bibitem{lei4dct2019}
    Y.~Lei, Y.~Fu, J.~Harms, T.~Wang, W.~J. Curran, T.~Liu, K.~Higgins, and
      X.~Yang, ``{4D-CT} deformable image registration using an unsupervised deep
      convolutional neural network,'' in \emph{Artificial {{Intelligence}} in
      {{Radiation Therapy}}}, 2019, pp. 26--33.
    
    \bibitem{yooSsEMnet2017}
    I.~Yoo, D.~G. Hildebrand, W.~F. Tobin, W.-C.~A. Lee, and W.-K. Jeong,
      ``{{ssEMnet}}: Serial-section electron microscopy image registration using a
      spatial transformer network with learned features,'' in \emph{Deep
      {{Learning}} in {{Medical Image Analysis}} and {{Multimodal Learning}} for
      {{Clinical Decision Support}}}, 2017, pp. 249--257.
    
    \bibitem{balakrishnanVoxelMorph2019}
    G.~Balakrishnan, A.~Zhao, M.~R. Sabuncu, J.~Guttag, and A.~V. Dalca,
      ``Voxelmorph: A learning framework for deformable medical image
      registration,'' \emph{IEEE Trans. Med. Imaging}, vol.~38, no.~8, pp.
      1788--1800, 2019.
    
    \bibitem{ganDeep2020}
    W.~Gan, Y.~Sun, C.~Eldeniz, J.~Liu, H.~An, and U.~S. Kamilov, ``Deep image
      reconstruction using unregistered measurements without groundtruth,''
      \emph{arXiv:2009.13986}, Sep. 2020.
    
    \bibitem{aggarwalMoDL2019}
    H.~K. Aggarwal, M.~P. Mani, and M.~Jacob, ``{{{MoDL}}:
      Model-based deep learning architecture for inverse problems},''
      \emph{{IEEE Trans. Med. Imaging}}, vol.~38, no.~2, pp.
      394--405, Feb. 2019.
    
    \bibitem{schlemperDeep2018}
    J.~Schlemper, J.~Caballero, J.~V. Hajnal, A.~N. Price, and D.~Rueckert,
      ``{A deep cascade of convolutional neural networks for
      dynamic {MR} image reconstruction},'' \emph{{IEEE
      Trans. Med. Imaging}}, vol.~37, no.~2, pp. 491--503, Feb. 2018.
    
    \bibitem{yangDeep2016}
    Y.~Yang, H.~Li, J.~Sun, and Z.~Xu, ``{Deep {{ADMM}}-net
      for compressive sensing {{MRI}}},'' in \emph{{Advances
      in Neural Information Processing Systems}}, 2016, p.~9.
    
    \bibitem{ronnebergerUnet2015}
    O.~Ronneberger, P.~Fischer, and T.~Brox, ``U-net: {{Convolutional}} networks
      for biomedical image segmentation,'' in \emph{Proc. Medical Image Computing
      and Computer-Assisted Intervention}, 2015, pp. 234--241.
    
    \bibitem{venkatakrishnan2013plug}
    S.~V. Venkatakrishnan, C.~A. Bouman, and B.~Wohlberg, ``Plug-and-play priors
      for model based reconstruction,'' in \emph{Proc. IEEE Global Conf. Signal
      Process. and Inf. Process. (GlobalSIP)}, 2013, pp. 945--948.
    
    \bibitem{romano2017little}
    Y.~Romano, M.~Elad, and P.~Milanfar, ``The little engine that could:
      Regularization by denoising {(RED)},'' \emph{SIAM J. Imaging Sci.}, vol.~10,
      no.~4, pp. 1804--1844, 2017.
    
    \bibitem{sreehari2016plug}
    S.~Sreehari, S.~V. Venkatakrishnan, B.~Wohlberg, G.~T. Buzzard, L.~F. Drummy,
      J.~P. Simmons, and C.~A. Bouman, ``Plug-and-play priors for bright field
      electron tomography and sparse interpolation,'' \emph{IEEE Trans. Comp.
      Imag.}, vol.~2, no. LA-UR-15-28750, 2016.
    
    \bibitem{zhang2017learning}
    K.~Zhang, W.~Zuo, S.~Gu, and L.~Zhang, ``Learning deep {CNN} denoiser prior for
      image restoration,'' in \emph{Proc. IEEE Conf. Computer Vision and Pattern
      Recognition}, 2017, pp. 3929--3938.
    
    \bibitem{sun2019regularized}
    Y.~Sun, S.~Xu, Y.~Li, L.~Tian, B.~Wohlberg, and U.~S. Kamilov, ``Regularized
      fourier ptychography using an online plug-and-play algorithm,'' in
      \emph{Proc. IEEE Int. Conf. Acoustics, Speech and Signal Process.
      (ICASSP)}.\hskip 1em plus 0.5em minus 0.4em\relax IEEE, 2019, pp. 7665--7669.
    
    \bibitem{zhang2019deep}
    K.~Zhang, W.~Zuo, and L.~Zhang, ``Deep plug-and-play super-resolution for
      arbitrary blur kernels,'' in \emph{Proc. IEEE Conf. Computer Vision and
      Pattern Recognition}, 2019, pp. 1671--1681.
    
    \bibitem{ahmad2020plug}
    R.~Ahmad, C.~A. Bouman, G.~T. Buzzard, S.~Chan, S.~Liu, E.~T. Reehorst, and
      P.~Schniter, ``Plug-and-play methods for magnetic resonance imaging: Using
      denoisers for image recovery,'' \emph{IEEE Signal Process. Mag.}, vol.~37,
      no.~1, pp. 105--116, 2020.
    
    \bibitem{gregor2010learning}
    K.~Gregor and Y.~LeCun, ``Learning fast approximations of sparse coding,'' in
      \emph{Proc. Int. Conf. Machine Learning}, 2010, pp. 399--406.
    
    \bibitem{zhang2018ista}
    J.~Zhang and B.~Ghanem, ``{ISTA-N}et: Interpretable optimization-inspired deep
      network for image compressive sensing,'' in \emph{Proc. IEEE Conf. Computer
      Vision and Pattern Recognition}, 2018, pp. 1828--1837.
    
    \bibitem{chen2015learning}
    Y.~Chen, W.~Yu, and T.~Pock, ``On learning optimized reaction diffusion
      processes for effective image restoration,'' in \emph{Proc. IEEE Conf.
      Computer Vision and Pattern Recognition}, 2015, pp. 5261--5269.
    
    \bibitem{liu2021sgd}
    J.~Liu, Y.~Sun, W.~Gan, X.~Xu, B.~Wohlberg, and U.~S. Kamilov, ``{SGD-Net}:
      Efficient model-based deep learning with theoretical guarantees,'' \emph{IEEE
      Trans. Comput. Imag.}, 2021.
    
    \bibitem{yaman2020self:mrm}
    B.~Yaman, S.~A.~H. Hosseini, S.~Moeller, J.~Ellermann, K.~U{\u{g}}urbil, and
      M.~Ak{\c{c}}akaya, ``Self-supervised learning of physics-guided
      reconstruction neural networks without fully sampled reference data,''
      \emph{Magn. Reson. Med.}, vol.~84, no.~6, pp. 3172--3191, 2020.
    
    \bibitem{yaman2021zeroshot}
    B.~Yaman, S.~A.~H. Hosseini, and M.~Akçakaya, ``Zero-shot self-supervised
      learning for {MRI} reconstruction,'' \emph{arXiv:2102.07737}, 2021.
    
    \bibitem{batsonNoise2self2019}
    J.~Batson and L.~Royer, ``{Noise2Self}: {{Blind}} denoising by
      self-supervision,'' in \emph{Proc. Int. Conf. Machine Learning}, 2019, pp.
      524--533.
    
    \bibitem{krullProbabilistic2020}
    A.~Krull, T.~Vi{\v c}ar, M.~Prakash, M.~Lalit, and F.~Jug, ``Probabilistic
      {{Noise2Void}}: {{Unsupervised}} content-aware denoising,'' \emph{Frontiers
      in Computer Science}, vol.~2, Feb. 2020.
    
    \bibitem{soltanayevTraining2018}
    S.~Soltanayev and S.~Y. Chun, ``{Training deep learning
      based denoisers without ground truth data},'' in
      \emph{{Advances in Neural Information Processing
      Systems}}, 2018, p.~11.
    
    \bibitem{quanSelf2Self2020}
    Y.~Quan, M.~Chen, T.~Pang, and H.~Ji, ``{{Self2Self with dropout}}: Learning
      self-supervised denoising from single image,'' in \emph{Proc. IEEE Conf.
      Computer Vision and Pattern Recognition}, Jun. 2020, pp. 1887--1895.
    
    \bibitem{ulyanov2018deep}
    D.~Ulyanov, A.~Vedaldi, and V.~Lempitsky, ``Deep image prior,'' in \emph{Proc.
      IEEE Conf. Computer Vision and Pattern Recognition}, 2018, pp. 9446--9454.
    
    \bibitem{liu2019image}
    J.~Liu, Y.~Sun, X.~Xu, and U.~S. Kamilov, ``Image restoration using total
      variation regularized deep image prior,'' in \emph{Proc. IEEE Int. Conf.
      Acoustics, Speech and Signal Process. (ICASSP)}.\hskip 1em plus 0.5em minus
      0.4em\relax IEEE, 2019, pp. 7715--7719.
    
    \bibitem{yoo2021time}
    J.~Yoo, K.~H. Jin, H.~Gupta, J.~Yerly, M.~Stuber, and M.~Unser,
      ``Time-dependent deep image prior for dynamic {MRI},'' \emph{IEEE Trans. Med.
      Imaging}, 2021.
    
    \bibitem{mataev2019deepred}
    G.~Mataev, P.~Milanfar, and M.~Elad, ``{DeepRED}: Deep image prior powered by
      {RED},'' in \emph{Proc. IEEE Int. Conf. Comput. Vis. Workshops}, 2019, pp.
      0--0.
    
    \bibitem{yang2012nonrigid}
    X.~Yang, P.~Ghafourian, P.~Sharma, K.~Salman, D.~Martin, and B.~Fei, ``Nonrigid
      registration and classification of the kidneys in {3D} dynamic contrast
      enhanced {(DCE) MR} images,'' in \emph{Proc. SPIE Int. Soc. Opt. Eng.}, vol.
      8314, 2012, p. 83140B.
    
    \bibitem{han2008atlas}
    X.~Han, M.~S. Hoogeman, P.~C. Levendag, L.~S. Hibbard, D.~N. Teguh, P.~Voet,
      A.~C. Cowen, and T.~K. Wolf, ``Atlas-based auto-segmentation of head and neck
      {CT} images,'' in \emph{Proc. Medical Image Computing and Computer-Assisted
      Intervention}, 2008, pp. 434--441.
    
    \bibitem{fu2017automatic}
    Y.~Fu, S.~Liu, H.~H. Li, and D.~Yang, ``Automatic and hierarchical segmentation
      of the human skeleton in {CT} images,'' \emph{Phys. Med. Biol.}, vol.~62,
      no.~7, p. 2812, 2017.
    
    \bibitem{bajcsyMultiresolution1989}
    R.~Bajcsy and S.~Kova{\v c}i{\v c}, ``Multiresolution elastic matching,''
      \emph{Comput. Vis., Graph., and Image Process.}, vol.~46, no.~1, pp. 1--21,
      1989.
    
    \bibitem{jaderbergSpatial2015}
    M.~Jaderberg, K.~Simonyan, and A.~Zisserman, ``Spatial transformer networks,''
      in \emph{Advances in Neural Information Processing Systems}, vol.~2, 2015,
      pp. 2017--2025.
    
    \bibitem{fengXD2016}
    L.~Feng, L.~Axel, H.~Chandarana, K.~T. Block, D.~K. Sodickson, and R.~Otazo,
      ``{{XD}}-{{GRASP}}: Golden-angle radial {{MRI}} with reconstruction of extra
      motion-state dimensions using compressed sensing,'' \emph{Magn. Reson. Med.},
      vol.~75, no.~2, pp. 775--788, Feb. 2016.
    
    \bibitem{fengHighly2013}
    L.~Feng, M.~B. Srichai, R.~P. Lim, A.~Harrison, W.~King, G.~Adluru, E.~V.~R.
      Dibella, D.~K. Sodickson, R.~Otazo, and D.~Kim, ``Highly accelerated
      real-time cardiac cine {{MRI}} using k-t {{SPARSE}}-{{SENSE}},'' \emph{Magn.
      Reson. Med}, vol.~70, no.~1, pp. 64--74, Jul. 2013.
    
    \bibitem{otazoCombination2010}
    R.~Otazo, D.~Kim, L.~Axel, and D.~K. Sodickson, ``Combination of compressed
      sensing and parallel imaging for highly accelerated first-pass cardiac
      perfusion {MRI},'' \emph{Magn. Reson. Med.}, vol.~64, no.~3, pp. 767--776,
      Sep. 2010.
    
    \bibitem{usmanMotion2013}
    M.~Usman, D.~Atkinson, F.~Odille, C.~Kolbitsch, G.~Vaillant, T.~Schaeffter,
      P.~G. Batchelor, and C.~Prieto, ``Motion corrected compressed sensing for
      free-breathing dynamic cardiac {{MRI}}: Motion {{Corrected Compressed
      Sensing}},'' \emph{Magn. Reson. Med.}, vol.~70, no.~2, pp. 504--516, Aug.
      2013.
    
    \bibitem{cruzHighly2017}
    G.~Cruz, D.~Atkinson, M.~Henningsson, R.~M. Botnar, and C.~Prieto, ``Highly
      efficient nonrigid motion-corrected {{3D}} whole-heart coronary vessel wall
      imaging,'' \emph{Magn. Reson. Med.}, vol.~77, no.~5, pp. 1894--1908, May
      2017.
    
    \bibitem{bustin3D2020}
    A.~Bustin, I.~Rashid, G.~Cruz, R.~Hajhosseiny, T.~Correia, R.~Neji, R.~Rajani,
      T.~F. Ismail, R.~M. Botnar, and C.~Prieto, ``{{3D}} whole-heart isotropic
      sub-millimeter resolution coronary magnetic resonance angiography with
      non-rigid motion-compensated {{PROST}},'' \emph{J. Cardiovasc. Magn. Reson.},
      vol.~22, no.~1, p.~24, Dec. 2020.
    
    \bibitem{blumeJoint2010}
    M.~Blume, A.~{Martinez-Moller}, A.~Keil, N.~Navab, and M.~Rafecas, ``Joint
      reconstruction of image and motion in gated positron emission tomography,''
      \emph{IEEE Trans. Med. Imaging}, vol.~29, no.~11, pp. 1892--1906, Nov. 2010.
    
    \bibitem{odilleJoint2016}
    F.~Odille, A.~Menini, J.-M. Escanye, P.-A. Vuissoz, P.-Y. Marie, M.~Beaumont,
      and J.~Felblinger, ``Joint reconstruction of multiple images and motion in
      {MRI}: Application to free-breathing myocardial {T2} quantification,''
      \emph{IEEE Trans. Med. Imaging}, vol.~35, no.~1, pp. 197--207, Jan. 2016.
    
    \bibitem{coronaMultitasking2019}
    V.~Corona, A.~I. {Aviles-Rivero}, N.~Debroux, M.~Graves, C.~Le~Guyader, C.-B.
      Sch{\"o}nlieb, and G.~Williams, ``Multi-tasking to correct:
      motion-compensated {MRI} via joint reconstruction and registration,'' in
      \emph{Scale {{Space}} and {{Variational Methods}} in {{Computer Vision}}},
      J.~Lellmann, M.~Burger, and J.~Modersitzki, Eds., vol. 11603.\hskip 1em plus
      0.5em minus 0.4em\relax {Cham}: {Springer International Publishing}, 2019,
      pp. 263--274.
    
    \bibitem{munozSelfsupervised2022}
    C.~Munoz, H.~Qi, G.~Cruz, T.~K{\"u}stner, R.~M. Botnar, and C.~Prieto,
      ``Self-supervised learning-based diffeomorphic non-rigid motion estimation
      for fast motion-compensated coronary {{MR}} angiography,'' \emph{Magn. Reson.
      Imaging}, vol.~85, pp. 10--18, Jan. 2022.
    
    \bibitem{qiEnd2021}
    H.~Qi, R.~Hajhosseiny, G.~Cruz, T.~Kuestner, K.~Kunze, R.~Neji, R.~Botnar, and
      C.~Prieto, ``End-to-end deep learning nonrigid motion-corrected
      reconstruction for highly accelerated free-breathing coronary {{MRA}},''
      \emph{Magn. Reson. Med.}, vol.~86, no.~4, pp. 1983--1996, Oct. 2021.
    
    \bibitem{kingmaAdam2014}
    D.~P. Kingma and J.~Ba, ``Adam: {{A}} method for stochastic optimization,''
      \emph{arXiv:1412.6980}, 2014.
    
    \bibitem{limEnhanced2017}
    B.~Lim, S.~Son, H.~Kim, S.~Nah, and K.~Mu~Lee, ``Enhanced deep residual
      networks for single image super-resolution,'' in \emph{Proc. IEEE Conf.
      Computer Vision and Pattern Recognition Workshops}, 2017.
    
    \bibitem{avantsSymmetric2008}
    B.~B. Avants, C.~L. Epstein, M.~Grossman, and J.~C. Gee, ``Symmetric
      diffeomorphic image registration with cross-correlation: Evaluating automated
      labeling of elderly and neurodegenerative brain,'' \emph{Med. Image Anal.},
      vol.~12, no.~1, pp. 26--41, 2008.
    
    \bibitem{sokootiNonrigid2017}
    H.~Sokooti, B.~De~Vos, F.~Berendsen, B.~P. Lelieveldt, I.~I{\v s}gum, and
      M.~Staring, ``Nonrigid image registration using multi-scale {{3D}}
      convolutional neural networks,'' in \emph{Proc. Medical Image Computing and
      Computer-Assisted Intervention}, 2017, pp. 232--239.
    
    \bibitem{eldenizConsistentlyAcquired2018}
    C.~Eldeniz, T.~Fraum, A.~Salter, Y.~Chen, H.~. Gach, P.~Parikh, K.~Fowler, and
      H.~An, ``Consistently-acquired projections for tuned and robust
      estimation--{A} self-navigated respiratory motion correction approach,''
      \emph{Invest. Radiol.}, vol.~53, no.~5, p. 293, 2018.
    
    \bibitem{avants2009advanced}
    B.~B. Avants, N.~Tustison, and G.~Song, ``Advanced normalization tools
      (ants),'' \emph{Insight j}, vol.~2, no. 365, pp. 1--35, 2009.
    
    \bibitem{renFaster2017}
    S.~Ren, K.~He, R.~Girshick, and J.~Sun, ``Faster {R}-{CNN}: {{Towards}}
      real-time object detection with region proposal networks,'' \emph{IEEE Trans.
      Pattern Anal. Mach. Intell.}, vol.~39, no.~6, pp. 1137--1149, 2017.
    
    \bibitem{lamontagneOASIS32019}
    P.~J. LaMontagne, T.~L. Benzinger, J.~C. Morris, S.~Keefe, R.~Hornbeck,
      C.~Xiong, E.~Grant, J.~Hassenstab, K.~Moulder, A.~Vlassenko \emph{et~al.},
      ``{{OASIS}}-3: Longitudinal neuroimaging, clinical, and cognitive dataset for
      normal aging and {{Alzheimer}} disease,'' \emph{MedRxiv 2019.12.13.19014902},
      2019.
    
    \bibitem{muckley:20:tah}
    M.~J. Muckley, R.~Stern, T.~Murrell, and F.~Knoll, ``{TorchKbNufft}: A
      high-level, hardware-agnostic non-uniform fast {Fourier} transform,'' in
      \emph{ISMRM Workshop on Data Sampling \& Image Reconstruction}, 2020.
    
    \bibitem{feng2014golden}
    L.~Feng, R.~Grimm, K.~T. Block, H.~Chandarana, S.~Kim, J.~Xu, L.~Axel, D.~K.
      Sodickson, and R.~Otazo, ``Golden-angle radial sparse parallel {MRI}:
      combination of compressed sensing, parallel imaging, and golden-angle radial
      sampling for fast and flexible dynamic volumetric {MRI},'' \emph{Magn. Reson.
      Med.}, vol.~72, no.~3, pp. 707--717, 2014.
    
    \bibitem{gordon2014dynamic}
    Y.~Gordon, S.~Partovi, M.~M{\"u}ller-Eschner, E.~Amarteifio, T.~B{\"a}uerle,
      M.-A. Weber, H.~Kauczor, and F.~Rengier, ``Dynamic contrast-enhanced magnetic
      resonance imaging: fundamentals and application to the evaluation of the
      peripheral perfusion,'' \emph{Cardiovascular diagnosis and therapy}, vol.~4,
      no.~2, p. 147, 2014.
    
    \bibitem{kamilovOptical2016}
    U.~S. Kamilov, I.~N. Papadopoulos, M.~H. Shoreh, A.~Goy, C.~Vonesch, M.~Unser,
      and D.~Psaltis, ``Optical tomographic image reconstruction based on beam
      propagation and sparse regularization,'' \emph{IEEE Trans. Comput. Imag.},
      vol.~2, no.~1, pp. 59--70, 2016.
    
    \end{thebibliography}


\end{document}